\documentclass[twocolumn,showpacs,preprintnumbers,superscriptaddress,prb,epsf]{revtex4}
\pdfoutput=1
\usepackage{graphics, bm}
\usepackage{psfrag}
\usepackage{amsmath}
\usepackage{amssymb}
\usepackage{epsfig}
\usepackage{feynmf}
\usepackage{afterpage}
\newcommand{\beq}    {\begin{equation}}
\newcommand{\enq}    {\end{equation}}
\newcommand{\ceq}[1] {Eq.(\ref{#1})}
\newcommand{\kk}     {{\rm \bf k }}
\newcommand{\pp}     {{\rm \bf p }}
\newcommand{\qq}     {{\rm \bf q }}
\newcommand{\Ss}     {{\rm \bf S }}
\newcommand{\dg}     {{\Delta\Gamma}}
\newcommand{\om}     {\omega_n}
\newcommand{\omn}    {\omega_n}

\newcommand{\num}    {\nu_m}

\newcommand{\numh}    {\hat\nu_m}

\newcommand{\omo}    {\omega_0}

\newcommand{\Om}     {\Omega_m}

\newcommand{\df}     {\equiv}
\newcommand{\ph}     {\varphi}

\newcommand{\sign}   {{\rm sign}}

\newcommand{\ecr}    {c^\dagger_{\kk\sigma}}
\newcommand{\ean}    {c_{\kk\sigma}}

\begin{document}

\title{Vertex corrections of impurity scattering at a ferromagnetic quantum critical point}

\author{Enrico Rossi}
\altaffiliation{Current address: Condensed Matter Theory Center,
Department of Physics, University of Maryland, College Park, MD
20742, USA.} \affiliation{Department of Physics, University of
Illinois at Chicago, Chicago, IL 60607, USA}
\author{Dirk K  Morr}
\affiliation{Department of Physics, University of Illinois at
Chicago, Chicago, IL 60607, USA} \affiliation{James Franck
Institute, University of Chicago, Chicago, Illinois 60637, USA}

\date{\today}


\begin{abstract}

We study the renormalization of a non-magnetic impurity's scattering
potential due to the presence of a massless collective spin mode at
a ferromagnetic quantum critical point. To this end, we compute the
lowest order vertex corrections in two- and three-dimensional
systems, for arbitrary scattering angle and frequency of the
scattered fermions, as well as band curvature. We show that only for
backward scattering in $D=2$ does the lowest order vertex correction
diverge logarithmically in the zero frequency limit. In all other
cases, the vertex corrections approach a finite (albeit possibly
large) value for $\omega \rightarrow 0$. We demonstrate that vertex
corrections are strongly suppressed with increasing curvature of the
fermionic bands. Moreover, we show how the frequency dependence of
vertex corrections varies with the scattering angle. We also discuss
the form of higher order ladder vertex corrections and show that
they can be classified according to the zero-frequency limit of the
lowest order vertex correction. We show that even in those cases
where the latter is finite, summing up an infinite series of ladder
vertex diagrams can lead to a strong enhancement (or divergence) of
the impurity's scattering potential. Finally, we suggest that the
combined frequency and angular dependence of vertex corrections
might be experimentally observable via a combination of frequency
dependent and local measurements, such as scanning tunneling
spectroscopy on ordered impurity structures, or measurements of the
frequency dependent optical conductivity.

\end{abstract}




\maketitle


\section{Introduction}

Understanding the complex physical properties of materials near a
quantum critical point (QCP) is one of the most important open
problems in condensed matter physics (for a recent review, see
\cite{Loe07} and references therein). An important piece of this
puzzle is the question of how soft collective fluctuations
associated with the proximity of the QCP affect the scattering
potential of non-magnetic impurities \cite{Miyake:2001,Kim:2003}.
The answer to this question is of great significance for a series of
experimental probes, ranging from measurements of the residual
resistivity \cite{Paul:2005,Belitz:2000} and optical conductivity to
that of the local density of states near impurities. Indeed, it was
recently argued by Kim and Millis~\cite{Kim:2003} that the transport
anomalies observed at the metamagnetic transition of ${\rm
Sr_3Ru_2O_7}$ \cite{Perry:2001,Grigera:2001} arise from a diverging
renormalization of the impurities' scattering potential.

In this article we study the renormalization of a non-magnetic
impurity's scattering potential due to the presence of a massless
collective spin mode at a ferromagnetic quantum critical point
(FMQCP). To this end, we compute both analytically and numerically,
the lowest order vertex corrections in two- and three-dimensional
systems, for arbitrary scattering angle and frequency of the
scattered electrons, as well as curvature of the fermionic band. We
limit ourselves to the case of a vanishing impurity density such
that the nature of the QCP is not altered by the presence of the
impurities. Moreover, since the renormalization of the impurity
potential by a collective mode effectively enlarges the size of the
impurity in real space, the limit of vanishing impurity density also
guarantees that quantum interference effects between impurities can
be omitted. The interaction of a soft collective spin mode with the
fermionic degrees of freedom in the vicinity of a FMQCP was studied
previously \cite{Rou01,Wang01,Chu03,Dze04,Chu05} (for related work
in the context of gauge theories see
\cite{Lee89,Blo93,Khv94,Nay94,Alt95}). At the FMQCP, the same
collective spin mode that renormalizes the impurity's scattering
potential yields self-energy corrections to the electronic Greens
functions that render the latter non-Fermi-liquid (NFL) like
\cite{Lee89,Alt95}. In order to investigate how the NFL nature of
the fermions affect the vertex corrections, we contrast them with
those obtained using a Fermi-liquid (FL) form of the fermionic
propagators. By doing so, we make contact with two earlier studies
of vertex corrections near a FMQCP. Specifically, Miyake {\it et
al.} \cite{Miyake:2001}, using a FL form of the fermionic
propagators, studied vertex corrections for forward scattering near
a FMQCP in $D=3$, while Kim and Millis \cite{Kim:2003} investigated
the vertex renormalization for backward scattering in $D=2$ at zero
frequency.

We obtain a number of interesting results. First, we find that only
for backward scattering in $D=2$ does the lowest order vertex
correction diverge in the limit of vanishing frequency. This
divergence is logarithmic both for the NFL and the FL case. In all
other cases, the vertex corrections approach a finite (albeit
possibly large) value for $\omega \rightarrow 0$, with the overall
scale for vertex corrections being smaller in the NFL than in the FL
case. This result is in disagreement with the earlier finding of
Miyake {\it et al.}~\cite{Miyake:2001} who reported a logarithmic
divergence in frequency for forward scattering in $D=3$.  Second,
vertex corrections are in general strongly suppressed with
increasing curvature of the fermionic bands. For backward
scattering, this suppression obeys a qualitatively different
functional form in the NFL and FL cases. Third, we identify the
combined dependence of the vertex corrections on scattering angle,
frequency, and band curvature. Fourth, we discuss the form of higher
order ladder vertex corrections and show that they can be classified
according to the zero-frequency limit of the lowest order vertex
correction. We show that even in those cases where the latter is
finite, summing up an infinite series of ladder vertex diagrams can
lead to a strong enhancement (or divergence) of the impurity's
scattering potential. Finally, we suggest that the combined
frequency and angular dependence of vertex corrections might be
experimentally observable via a combination of frequency dependent
and local measurements, such as scanning tunneling spectroscopy
(STS) on ordered impurity structures, or measurements of the
frequency dependent optical conductivity.

While we consider below a FMQCP with a ${\bf q}=0$ ordering
wavevector, the question of whether such a QCP is actually realized
in nature has attracted significant interest over the last few
years. In the standard theoretical approaches
\cite{Hertz:1976,Millis:1993} such a QCP is described by integrating
out the electronic degrees of freedom and deriving a low energy
Landau-Ginzburg effective action in terms of the order parameter
field. A number of recent studies
\cite{Belitz:1997,Coleman:2001,Abanov:2003,Chubukov:2003} (for
reviews see \cite{Abanov:2003,Belitz:2002,Belitz:2005,Rech:2006})
have shed some doubt on the validity of this approach since they
argued that for a system with an SU(2) (Heisenberg) spin symmetry,
the momentum dependence of the static spin susceptibility acquires
negative non-analytic corrections. These corrections could either
lead to a first-order transition in the ferromagnetic state, or
could give rise to an incommensurate order with non-zero ordering
wavevector. In contrast, in systems with an Ising spin symmetry,
these corrections are absent, and a $({\bf q}=0)$-FMQCP can in
general occur. A candidate system for the latter case is ${\rm
Sr_3Ru_2O_7}$, which exhibits a metamagnetic transition. While this
transition is in general first order in nature, it can be tuned to a
critical end point which is located close to $T=0$, thus
representing a (quasi) quantum critical point. At the same time, the
externally applied field renders the spin symmetry Ising-like. In
order to keep the discussion of our results as general as possible,
we explicitly show below how the spin symmetry of a system,
Heisenberg-,  XY- or Ising-like, affects the form of the vertex
corrections.

The rest of the paper is organized as follow. In
Sec.~\ref{sec:self_energy} we give a brief derivation of the bosonic
and fermionic propagators at the FMQCP in $D=2$ and $3$. In
Sec.~\ref{sec:vc} we present the general expression for the lowest
order vertex correction, $\dg$, that we employ for our numerical and
analytical calculations.  In Secs.~\ref{sec:vc_2D} and
\ref{sec:vc_3D} we present the results for vertex corrections in
$D=2$ and $D=3$, respectively, as a function of scattering angle,
frequency and band curvature. In Sec.~\ref{sec:HigherOrder} we
discuss the form of the higher order vertex corrections and their
effects on the renormalized scattering potential. Finally, in
Sec.~\ref{sec:conclusion} we summarize our findings and discuss
their implications for transport and STS experiments.

%
%
\section{Spin susceptibility and fermionic self energy}
\label{sec:self_energy}

Our starting point is the spin-fermion model (for details, see,
e.g., Ref.~\cite{Abanov:2003,Rech:2006}) whose Hamiltonian is given
by
\begin{align}
 H =& \sum_{\kk\sigma} \epsilon_\kk\ecr\ean  +
      \sum_q \chi_0(\qq)^{-1}\Ss_\qq\Ss_{-\qq} \nonumber \\
    & + g\sum_{\kk,\qq,\sigma,\sigma'} \ecr {\bf \tau}_{\sigma\sigma'} c_{\kk+\qq,\sigma'}
    \cdot\Ss_\qq \quad ,
 \label{eq:ham}
\end{align}
where $c^\dagger_{{\bf k},\sigma}$, ($c_{{\bf k},\sigma}$) is the
fermionic creation (annihilation) operator for an electron with
momentum ${\bf k}$ and spin $\sigma$, $\Ss_q$ are vector bosonic
operators, $g$ is the effective fermion-boson coupling, ${\bf
\tau}_{\sigma\sigma'}$ are the Pauli matrices, and
\begin{equation}
\chi_0(\qq) =  \frac{\chi_0}{\xi^{-2} + q^2}
\end{equation}
is the static propagator describing ferromagnetic fluctuations, with
$\xi$ being the magnetic correlation length. While, in general,
$\chi_0(\qq)$ could be obtained by integrating out the high-energy
fermions (for a more detailed discussion, see Ref.\cite{Rech:2006}),
it is commonly used as a phenomenological input for the model. The
above model describes the low-energy excitations of the system (such
that $|\qq|<W/v_F$ where $W$ is of the order of the fermionic
bandwidth and $v_F$ is the Fermi velocity), in which the dynamic
part of bosonic propagator is generated by the interaction with the
low-energy fermions. The full bosonic propagator is thus obtained
via the Dyson equation
\beq
 \chi^{-1}(\qq,i\Omega_m) = \chi^{-1}_0(\qq) - \Pi(\qq,i\Omega_m)
 \quad ,
 \label{eq:chi1}
\enq
where $\Pi(\qq,i\Omega_m)$ is the bosonic self-energy (the
polarization operator) which to lowest order in $g$ is given by
\begin{align}
 \Pi(\qq,i\Omega_m) =& -g^2 T \sum_n \int \frac{d^D k}{(2\pi)^d} \
     G_0(\kk,i\omn) \nonumber \\
    & \times G_0(\kk+\qq,i\omn+i\Om) \quad ,
\end{align}
where $G_0(\kk,i\omn)=(i\omn -\epsilon_\kk)^{-1}$ is the bare
Green's function for the fermions, $\omn$ ($\Omega_m$) is the
fermionic (bosonic) Matsubara frequency, and $D$ is the
dimensionality of the system. In the remainder of the paper, we set
$\xi\to\infty$ and study the form of vertex corrections for impurity
scattering at the FMQCP.

\begin{figure}
 \begin{center}
  \includegraphics*[width=8cm]{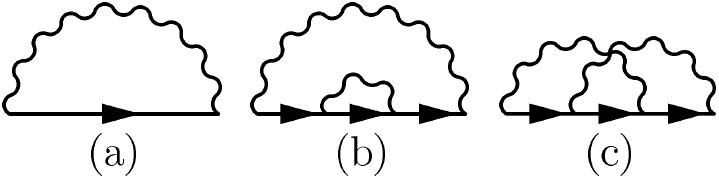}
  \caption{(a) Diagram for the lowest order self energy correction.
           (b) Example of rainbow diagram.
           (c) Example of neglected higher order corrections.
          }\label{fig:self_en_diag}
 \end{center}
\end{figure}

In the limit $T \to 0$, the polarization bubble can easily be
evaluated (see, for example, \cite{Abanov:2003,Rech:2006}). After
expanding the fermionic dispersion as \beq
  \epsilon_\kk \approx v_F(\kk) k_\parallel + r(\kk) (k_\parallel^2 +
  k_\perp^2) \quad ,
  \label{eq:disp}
\enq where $v_F(\kk)$ and $r(\kk)$ are the local Fermi velocity and
band curvature, respectively, and $k_\parallel$ ($k_\perp$) is the
component of $\kk$ parallel (perpendicular) to the normal of the
Fermi surface, and keeping only terms in $G_0$ up to linear order in
$\qq$, one obtains for an isotropic system (i.e., for constant $v_F$
and $r$) in $D=2$
\beq
 \Pi(\qq,i\Omega_m) = -2 g^2 N_0 \frac{|\Om|}{\sqrt{\Om^2 + (v_F
 q)^2}} \quad ,
 \label{eq:Pi2D}
\enq
while for $D=3$ one has
\beq
 \Pi(\qq,i\Omega_m) = i\frac{g^2 N_0 |\Om|}{v_F q}
              \ln\left(\frac{1-i\frac{\Om}{v_F q}}{-1-i\frac{\Om}{v_F
              q}}\right) \quad ,
 \label{eq:Pi3D}
\enq
where $q=|{\bf q}|$, and $N_0$ is the density of states. We assume
$r q/v_F \ll 1$ such that the local curvature $r$ of the fermionic
dispersion leads to only subleading corrections to $\Pi$, which can
be neglected \cite{Rech:2006}. In the limit $\Om\ll v_F q$ we can
simplify the above expressions, and obtain for $D=2$ from
\ceq{eq:Pi2D} \cite{Chu05}
\beq
 \Pi(\qq,i\Omega_m) = -2 g^2 N_0 \frac{|\Om|}{v_F q}
 \label{eq:Pi2D2}
\enq
and for $D=3$ from \ceq{eq:Pi3D} \beq
 \Pi(\qq,i\Omega_m) = -\pi g^2 N_0 \frac{|\Om|}{v_F q} \quad .
 \label{eq:Pi3D2}
\enq
In order to
simplify the notation, we introduce the Landau damping parameter
$\lambda$ such that for $D=2$ one has $\lambda\df 2\chi_0 g^2 N_0$
while for $D=3$ one obtains $\lambda\df \pi\chi_0 g^2 N_0$, the
difference being only a numerical factor. At the QCP, the full spin
susceptibility is then given by
\beq
 \chi(\qq,i\Om) = \frac{\chi_0}{q^2 + \lambda\frac{|\Om|}{v_F q}} \
 .
 \label{eq:chi2}
\enq
Using this form of $\chi$, we can now compute the lowest order
fermionic self energy correction, $\Sigma$, shown in
Fig.\ref{fig:self_en_diag}(a),
which for $T = 0$ is given by \beq
 \Sigma(\kk, i\om) = I_s \ g^2\int\frac{d^D p \ d\num}{(2\pi)^{D+1}}
                     \chi(\kk+\pp,i\om + i\num)G(\pp,i\num) \ ,
 \label{eq:sigma}
\enq
with $I_s=1, 2$ or 3 for a system with Ising, XY or Heisenberg spin
symmetry, respectively. In the limit $\lambda/k_F^2 \ll 1$, typical
bosonic momenta are much bigger than typical fermionic momenta
\cite{Rech:2006}, and one can decouple the momentum integration
parallel and perpendicular to the normal of the Fermi surface by
setting
\begin{equation}
\chi(p_\perp, p_\parallel,i \Omega_m) \approx \chi(p_\perp, 0,i
\Omega_m) \label{eq:decoup} \ .
\end{equation}
This yields
\begin{align}
 \Sigma(\kk, i\om) = &I_s g^2\int\frac{d^{D-1} p_\perp d\num}{(2\pi)^{D}}
                      \chi(p_\perp,i\num + i\om) \nonumber \\
                     & \times \int_{-\infty}^{\infty} \frac{d
                     p_\parallel}{2 \pi}
                     G(p_\parallel,i\num).
 \label{eq:sigma2}
\end{align}
Using $G(p_\parallel,i\omn)=1/(i\omn - v_F p_\parallel)$, one
obtains from \ceq{eq:sigma2} in $D=2$ (see Ref.\cite{Lee89,Alt95})
\beq
 \Sigma(\kk, i\om) = -i\omega_0^{1/3}|\om|^{2/3} {\rm sgn} (\om)
 \label{eq:sigma2D}
\enq
where \beq
 \omega_0 \df \frac{I^3_s }{3 \sqrt{3}(4\pi)^3}\frac{\lambda^2  }{v_F^2
 N_0^3} = \frac{I^3_s }{12 \sqrt{3}} \ \alpha^2 E_F\ ,
 \label{eq:omc2D}
\enq $\alpha=\lambda/k_F^2$, and $E_F=v_F k_F/2$ is the Fermi
energy. $\omega_0$ is interpreted as the upper frequency scale for
quantum critical behavior since for $\omega < \omega_0$ the fermions
possess a non-Fermi liquid character, while for $\omega
> \omega_0$ the Fermi liquid behavior is recovered. For $D=3$, one
has
\beq
 \Sigma(\qq, i\om) = -i \gamma \, \om \, \ln\left(1 + \frac{\Omega_c}{|\om|}\right)
 \label{eq:sigma3D}
\enq
with
\beq
 \gamma \df \frac{ I_s g^2\chi_0}{12 \pi^2 v_F}=I_s \frac{\lambda}{6 \pi k_F^2} ; \quad \Omega_c \df
 \frac{v_F\Lambda^3}{\lambda} \ .
 \label{eq:gamma}
\enq

Here, $\Lambda$ is a momentum cutoff of order $1/a_0$, where $a_0$
is the lattice constant. In $D=3$ the upper frequency scale for
quantum critical behavior is given by
\beq
 \omega_0\df\Omega_c e ^{-1/\gamma}.
 \label{eq:omo_3D}
\enq
As discussed above, this derivation of the self-energies only holds
in the limit $\lambda/k_F^2 \ll 1$, implying $\alpha \ll 1$ in $D=2$
and $\gamma \ll 1$ in $D=3$.

Finally, we briefly comment on the effect of higher order
self-energy diagrams. It turns out that rainbow diagrams, such as
the one shown in Fig.\ref{fig:self_en_diag}(b), vanish identically
as long as the lowest order self-energy in
Fig.\ref{fig:self_en_diag}(a) (which is an internal part of all
rainbow diagrams) is independent of $k_\parallel$; in this case the
poles of the internal Greens functions all lie in the same half of
the complex plane. In contrast, crossing diagrams such as the one
shown in Fig.\ref{fig:self_en_diag}(c), provide a renormalization of
the boson-fermion vertex, $g$, whose discussion is beyond the scope
of this article.


\section{Vertex corrections for Scattering of Non-Magnetic Impurities}
\label{sec:vc}
\begin{figure}
 \begin{center}
  \includegraphics*[width=8cm]{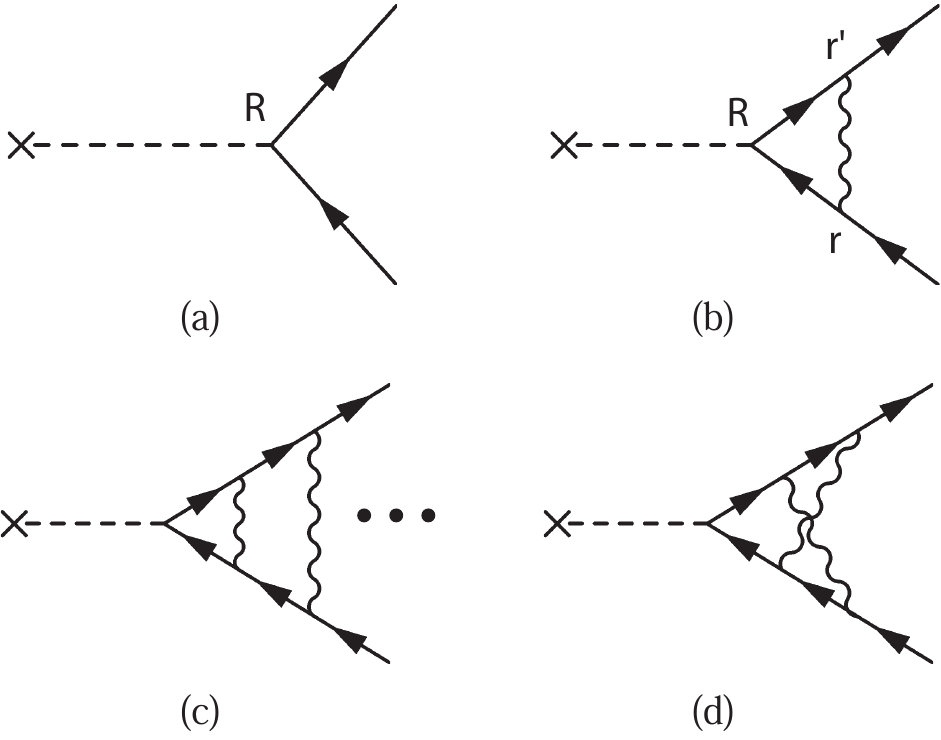}
  \caption{(a) Bare vertex.
           (b) Lowest order correction to the
           impurity scattering vertex due the exchange of
           soft magnetic fluctuations.
       (c) and (d) higher order diagrams: ladder (c), and crossed (d).
          }
  \label{fig:vcdiag}
 \end{center}
\end{figure}

We next consider the scattering of a single non-magnetic impurity,
located at ${\bf R}$. The bare scattering process is described by
the Hamiltonian ${\cal H}_{imp}=\sum_\sigma U_0 c^\dagger_\sigma
({\bf R}) c_\sigma ({\bf R})$ with bare scattering vertex $U_0$, and
represented by the diagram shown in Fig.~\ref{fig:vcdiag}(a). The
diagrams shown in Fig.~\ref{fig:vcdiag}(b)-(d) represent processes
involving the soft ferromagnetic mode that renormalize the
impurity's scattering potential. Note that these processes lead to a
renormalized scattering vertex which depends on two additional
spatial coordinates, ${\bf r}$ and ${\bf r}^\prime$ as shown in
Fig.~\ref{fig:vcdiag}(b), effectively leading to an increase in the
spatial size of the impurity.

We begin by studying the form of the lowest order vertex correction,
$- \dg$, shown in Fig.~\ref{fig:vcdiag}(b). Up to second order in
$g$, the full vertex, $U$, is then given by
\begin{equation}
U=U_0 \left(1+ \frac{\dg}{U_0} + ...\right) \quad .
\end{equation}
Fourier transformation of $\dg$ into momentum space yields at $T=0$
(where the fermionic Matsubara frequency $\num$ is now a continuous
variable)
\begin{align}
 \frac{\dg({\bf p},\kk,i\om)}{U_0} = &- I_s g^2
        \int \frac{d^D q}{(2\pi)^D} \int \frac{d\num}{2 \pi} \chi(\qq,i\num - i\om) \nonumber \\
       &G(\pp-\qq-\kk/2,i\num)G(\pp-\qq+\kk/2,i\num)
 \label{eq:dg1}
\end{align}
with momentum transfer $\kk$ at the impurity. In order to evaluate
$\dg$, we write $G$ in the general form
\beq
 G(\kk,i\num)=\frac{1}{Z(i\num) -\epsilon_k} \quad ,
 \label{eq:G_gen}
\enq
where $Z(i\num)=i\num$ in the Fermi-liquid regime, and $Z(i\num) =
i\num + \Sigma(i\num)$ in the non-Fermi liquid regime. Here,
$\Sigma$ for $D=2$ is given by \ceq{eq:sigma2D} and for $D=3$ by
\ceq{eq:sigma3D}. In what follows, we study $\dg$ for fermions near
the Fermi surface, and hence set $ \left|\pp - \kk/2\right|\sim p_F$
and $ \left|\pp + \kk/2\right|\sim p_F$. We can then expand the
dispersion entering the Greens functions in Eq.(\ref{eq:dg1}) as
\begin{eqnarray}
 \epsilon_{\pp-\qq-\kk/2}&\sim& -v_F q\cos\theta + r q^2 \nonumber
 \\
 \epsilon_{\pp-\qq+\kk/2}&\sim& -v_F q\cos(\theta +\varphi) + r q^2
 \quad ,
\end{eqnarray}
where $\ph$ is the scattering angle and $\theta$ is the angle
between $\pp - \kk/2$ and $\qq$. With these approximations and the
expression given in \ceq{eq:chi2} for the bosonic propagator, we
obtain in $D=2$
\begin{align}
 \frac{\dg}{U_0} = &- \frac{ I_s g^2\chi_0}{(2\pi)^{3}}
        \int_0^{q_{max}} d q\int_{-\omega_{max}}^{\omega_{max}} d\num
        \frac{q^2}{q^3 + \frac{\lambda}{v_F}|\num-\om|} \nonumber \\
       & \times \int_0^{2\pi} d\theta\nonumber
        \frac{1}{Z(\num) +v_F q\cos\theta - r q^2}\cdot    \nonumber \\
       & \times \frac{1}{Z(\num) +v_F q\cos(\theta +\varphi) - r
       q^2} \quad ,
 \label{eq:dg_2D}
\end{align}
while for $D=3$, one has
\begin{align}
 \frac{\dg}{U_0} = & - \frac{I_s g^2\chi_0}{(2\pi)^{3}}
        \int_0^{q_{max}} d q \int_{-\omega_{max}}^{\omega_{max}} d\num
        \frac{q^3}{q^3 + \frac{\lambda}{v_F}|\num-\om|} \nonumber \\
       &\times \int_{-1}^{1} d(\cos\theta)
    \frac{1}{Z(\num) +v_F q\cos\theta - r q^2}\cdot    \nonumber \\
       & \times \frac{1}{Z(\num) +v_F q\cos(\theta +\varphi) - r q^2} \quad .
 \label{eq:dg_3D}
\end{align}
Here, $q_{max}$ and $\omega_{max}$ are cut-offs in momentum and
frequency space, respectively. All numerical results presented below
for $\Delta \Gamma$ in $D=2$ are obtained from Eq.(\ref{eq:dg_2D})
with $Z(\num)=i\num$ for the FL case, and $Z(\num)=i \num + i
\omo^{1/3}|\num|^{2/3} {\rm sgn} (\num)$  for the NFL case
\cite{com2}. Moreover, after rescaling all frequencies with
$\omega_0$ and all momenta with $q_0=\omega_0/v_F$, one finds that
the above integral depends only on the dimensionless quantities
$\alpha=\lambda/k_F^2$ and
\begin{equation}
r_d=\frac{r q_0^2}{\omega_0} = \frac{3 \pi \sqrt{3}}{4} r N_0 \,
\alpha^2 \quad ,
\end{equation}
where $N_0$ is the density of states of the clean system. For the
numerical results in $D=2$ shown below, we use $q_{max}=5000 q_0$
and $\omega_{max}=5000 \omega_0$.

Similarly, the numerical results for $\dg$ in $D=3$ are obtained
from Eq.(\ref{eq:dg_3D}) with $Z(\num)= i \num$ for the FL case, and
$Z(\num)= i \num + i \gamma  \num \ln \left( 1+\Omega_c/|\num|
\right)$ for the NFL case. After rescaling of momentum and
frequency, the integral in Eq.(\ref{eq:dg_3D}) depends only on the
dimensionless quantities $\gamma$ and $r_d$, where the latter in
$D=3$ is given by
\begin{equation}
r_d= \frac{\pi}{k_F} r N_0 \left(\frac{\Lambda}{k_F} \right)^3
\frac{e^{-1/\gamma}}{\gamma} \quad .
\end{equation}
The respective values for $q_{max}$ and $\omega_{max}$ employed in
the numerical evaluation of Eq.(\ref{eq:dg_3D}) are given below.

In order to complement the numerical results for $\dg$ shown below,
we consider analytically the cases of forward ($\varphi=0$) and
backward ($ \varphi=\pi$) scattering. Employing the same momentum
decoupling [see Eq.(\ref{eq:decoup})] as was used for the
calculation of the fermionic self-energy, one obtains from
Eq.(\ref{eq:dg1})
\begin{eqnarray}
\frac{\dg}{U_0} = -\frac{I_s g^2}{(2 \pi)^{D+1}}
\int_{-\infty}^{\infty}
d\num \int_{-\infty}^{\infty} dq^{D-1}_\perp \chi(q_\perp,i\num - i\om) \nonumber \\
\times \int_{-\infty}^{\infty} dq_\parallel
\frac{1}{Z(i\num)-v_Fq_\parallel-rq_\perp^2}\frac{1}{Z(i\num) \mp
v_Fq_\parallel-rq_\perp^2} \ \ \label{eq:an}
\end{eqnarray}
where in the last line, the terms $- v_Fq_\parallel$ and $+
v_Fq_\parallel$ correspond to forward and backward scattering,
respectively. Eq.(\ref{eq:an}) is the starting point for the
analytical results presented below.


\section{Vertex Corrections in $D=2$}
\label{sec:vc_2D}
In this section we discuss the frequency and angular dependence of
$\Delta \Gamma$ in $D=2$ for both the FL and NFL cases. For the
numerical results presented in this section, we use for definiteness
$\alpha=0.1$ and $I_s=3$.

\subsection{Forward scattering}
%

%
\begin{figure}[!h]
 \begin{center}
  \includegraphics*[width=8cm]{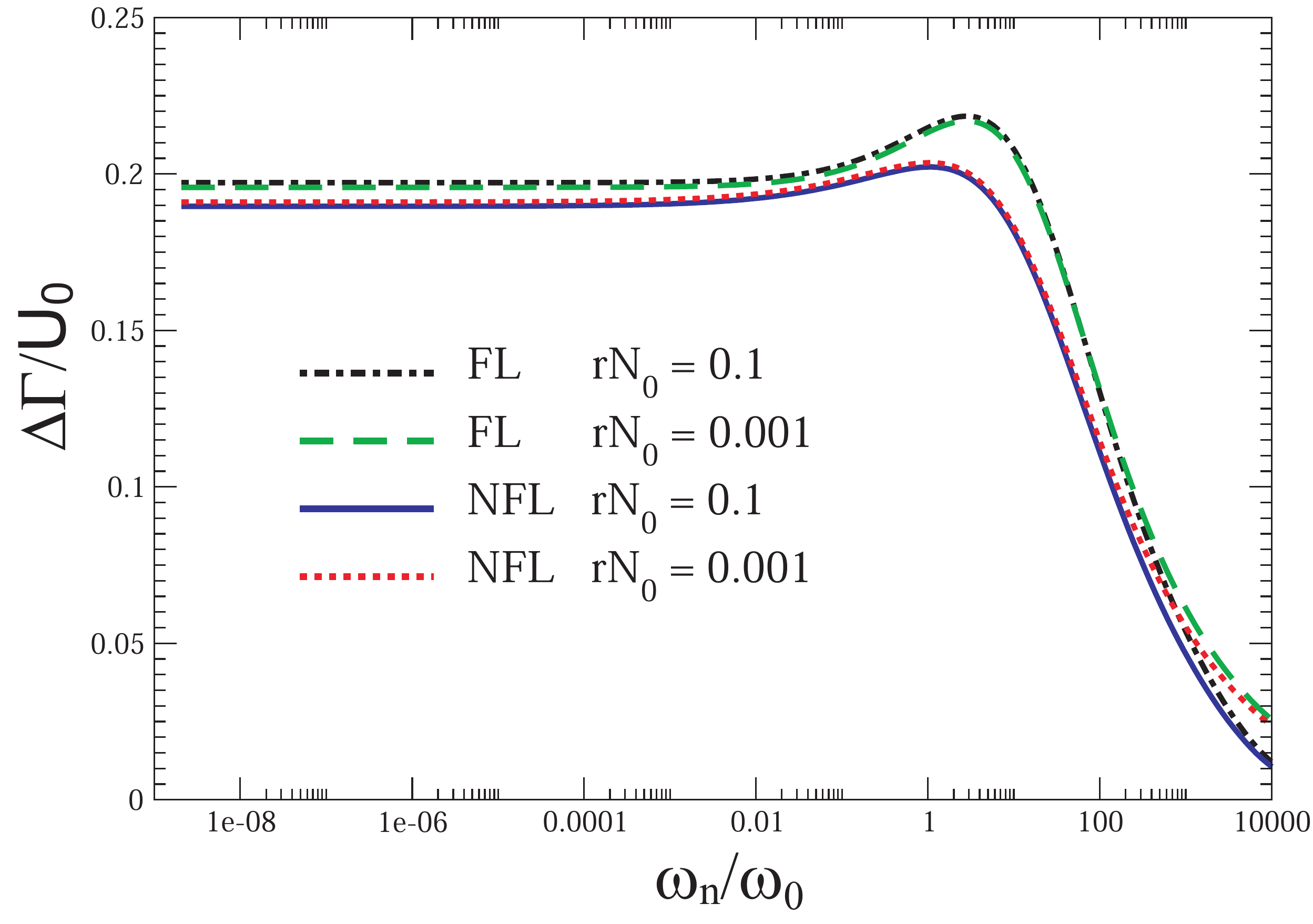}
  \caption{
           Vertex corrections for forward scattering in $D=2$
       in the FL and NFL cases for several values of the curvature $r$.
          }
  \label{fig:2D_forw}
 \end{center}
\end{figure}
In Fig.~\ref{fig:2D_forw} we present the numerical results obtained
from \ceq{eq:dg_2D} for forward scattering ($\varphi=0$) in the NFL
and FL cases for different values of the curvature, $r$. We find
that in both cases, $\dg$ approaches a finite value in the limit
$\omega_n \rightarrow 0$, consistent with the findings by Rech {\it
et al.}~\cite{Rech:2006}, and is almost frequency independent upto a
frequency of order $O(\omega_0)$ where it exhibits a weak maximum
before rapidly decreasing at larger frequencies. Moreover, $\Delta
\Gamma$ exhibits a very similar functional form for the NFL and FL
cases exhibiting only a small quantitative difference in the overall
scale. Note that in both cases, $\dg$ varies only weakly with
curvature $r$. In order to understand the similar frequency
dependence in the NFL and FL cases, we evaluate Eq.(\ref{eq:an})
analytically. Since $\Delta \Gamma$ depends only weakly on $r$, we
set for simplicity $r=0$. Moreover, since the poles of the Greens
functions in Eq.(\ref{eq:an}) lie in the same half of the complex
plane for $\varphi=0$, it is necessary to introduce a finite upper
cut-off, $q_{max}$, in the momentum integration in order to obtain a
non-zero result for $\Delta \Gamma$. Note that the numerical
evaluation of Eq.(\ref{eq:dg_2D}), in which the momentum decoupling,
Eq.(\ref{eq:decoup}), is not employed, yields a finite value for
$\Delta \Gamma$ even in the limit $q_{max} \rightarrow \infty$ due
to the branch cut contribution coming from $\chi$ \cite{Rech:2006}.
After performing the momentum integration in Eq.(\ref{eq:an}) and
rescaling $\numh\df \frac{\num}{v_F q_{max}}$ one obtains
\beq
 \frac{\dg}{U_0} = -\frac{2 I_s}{9\pi}\left(\frac{\omega_0}{v_F q_{max}}\right)^{1/3}
        \int_{-\infty}^\infty
        \frac{d \numh}{1 + Z^2(\numh)}\frac{1}{|\numh-{\hat
        \omega_n}|^{1/3}} \ .
 \label{eq:2D_forw_fl2}
\enq
In the limit ${\hat \omega}_0=\omega_0/ v_F q_{max} \ll 1$
considered here (for the numerical results in
Fig.~\ref{fig:2D_forw}, one has ${\hat \omega}_0=2 \times 10^{-4}$),
one obtains $|\numh+{\hat \omega}_0^{1/3} \numh^{2/3}| \approx
|\numh|$ for $\numh \gg {\hat \omega}_0$, implying that the term $[1
+ Z^2(\numh)]^{-1}$ in the integrand of Eq.(\ref{eq:2D_forw_fl2}),
is approximately the same for the FL and NFL cases, thus explaining
the small quantitative differences between these two cases.

%
%
\subsection{Backward scattering}

We begin by discussing the FL case, for which the analytical
derivation of $\dg$, starting from Eq.(\ref{eq:an}), is presented in
Appendix \ref{A1}. Since the final expression for $\dg$ given in
Eq.(\ref{eq:dgfinal}) is rather cumbersome, we fill focus here on
two limiting cases. In the low-frequency limit $\omega_n \ll
\omega_r$, the asymptotic behavior of $\dg$ is given by
\begin{equation}
 \frac{\dg}{U_0} = \frac{3}{16\pi } \frac{1}{N_0 r} \ln \left( \frac{\omega_n}{\omega_0}
 \right) - O\left( \omega_n^{1/3} \right)
 \label{eq:dg_2d_bw_lf}
 \end{equation}
 while for $\omega_n \gg \omega_r$,  one obtains
 \begin{align}
 \frac{\dg}{U_0} = & 2 \left( \frac{\omega_n}{\omega_0}
 \right)^{-1/3} - \sqrt{\frac{3}{2}} z^{-1/2} \left( \frac{\omega_n}{\omega_0}
 \right)^{-1/2} \nonumber \\
 & \quad - O\left[ \frac{\ln (\omega_n) }{\omega_n} \right]
 \label{eq:dg_2d_bw_hf}
\end{align}
with
\begin{equation}
z=\frac{\sqrt{3}}{4 \pi } \frac{1}{r N_0} \ . \label{eq:z}
\end{equation}
Here, $\omega_r=z^{-3}\omega_0 \sim r^3$ is the crossover frequency
between the high and low frequency limits which strongly varies with
the curvature of the Fermi surface. In the low-frequency limit,
$\dg$ exhibits a logarithmic divergence whose prefactor depends
inversely on the local curvature, $r$. In contrast, in the high
frequency limit, the leading order frequency dependence of $\dg$ is
algebraic with an $r$-independent prefactor.

\begin{figure}[!h]
 \begin{center}
  \includegraphics*[width=8.5cm]{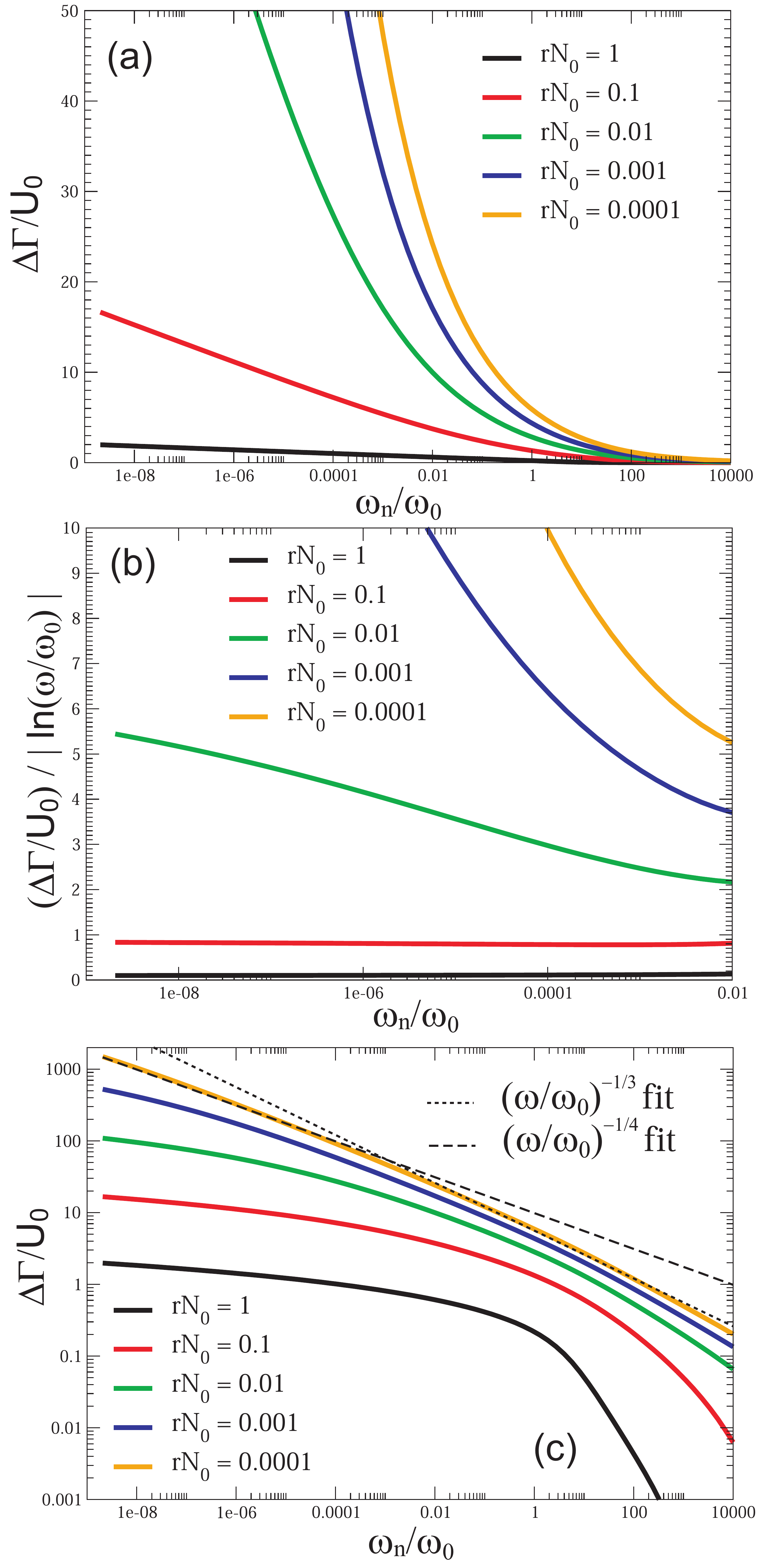}
  \caption{Backward scattering in 2D for the FL case: (a) $\Delta \Gamma/U_0$, (b)
  $(\Delta \Gamma/U_0)/|\ln \left(\omega_n/\omega_0\right)|$,
  (c) log-log plot of $\dg/U_0$ with fits to $(\omega_n/\omega_0)^{-1/3}$ (dotted line) and  $(\omega_n/\omega_0)^{-1/4}$ (dashed line).}
  \label{fig:2D_bw_FL}
 \end{center}
\end{figure}
In Fig.~\ref{fig:2D_bw_FL}(a), we present the frequency dependence
of $\Delta \Gamma$ for the FL case and backward scattering obtained
numerically from Eq.(\ref{eq:dg_2D}) for different values of
curvature $r$. In agreement with the analytical results in
Eq.(\ref{eq:dg_2d_bw_lf}), we find that $\dg$ diverges in the limit
$\omega_n \rightarrow 0$ and that its overall scale decreases with
increasing $r$. In order to extract the functional dependence of
$\dg$ in the low-frequency limit, we plot in
Fig.~\ref{fig:2D_bw_FL}(b) the ratio $R=(\dg/U_0) / |\ln
(\omega_n/\omega_0)|$. For $r N_0=1$ and $0.1$, this ratio is
constant in the low-frequency limit, implying a logarithmically
diverging $\dg$, in agreement with Eq.(\ref{eq:dg_2d_bw_lf}). For $r
N_0=0.01$, $R$ still possesses a substantial frequency dependence at
low frequencies, however, its second derivative, $d^2R/d
\ln^2(\omega)$, is negative, suggesting that $R$ approaches a
constant value for $\omega_n \rightarrow 0$. However, for $r
N_0=0.001$ and $0.0001$, we find that $R$ strongly increases with
decreasing $\omega_n$ upto the smallest frequencies that we can
access numerically. The problem in establishing a logarithmic
divergence of $\dg$ for small values of $r N_0$ arises from the fact
that this divergence emerges only in the limit $\omega \ll \omega_r
\sim r^3 \omega_0$. Since $\omega_r$ decreases strongly with
decreasing curvature $r$, it becomes increasingly difficult to
identify the logarithmic divergence over the numerically accessible
frequency range. For example, for $r N_0=0.0001$, one has
$\omega_r=3.82 \times 10^{-10} \omega_0$ which is smaller than the
smallest frequencies we can consider. The approach to a logarithmic
divergence at low frequencies, however, can be seen  by considering
a log-log plot of $\dg$, as shown in Fig.~\ref{fig:2D_bw_FL}(c). For
$r N_0=0.0001$ we find that at larger frequencies $\omega_n/\omega_0
\gtrsim 10^{-3}$, $\dg$ scales approximately as $\omega_n^{-1/3}$
while in the low-frequency range $10^{-9} \gtrsim \omega_n/\omega_0
\gtrsim 10^{-4}$, we have $\dg \sim \omega_n^{-1/4}$ [see straight
lines in Fig.~\ref{fig:2D_bw_FL}(c)]. The decrease in the exponent
of the algebraic dependence with decreasing frequency suggests that
$\dg$ eventually crosses over to a logarithmic form in the limit
$\omega_n \rightarrow 0$.

\begin{figure}[!h]
 \begin{center}
  \includegraphics*[width=8.5cm]{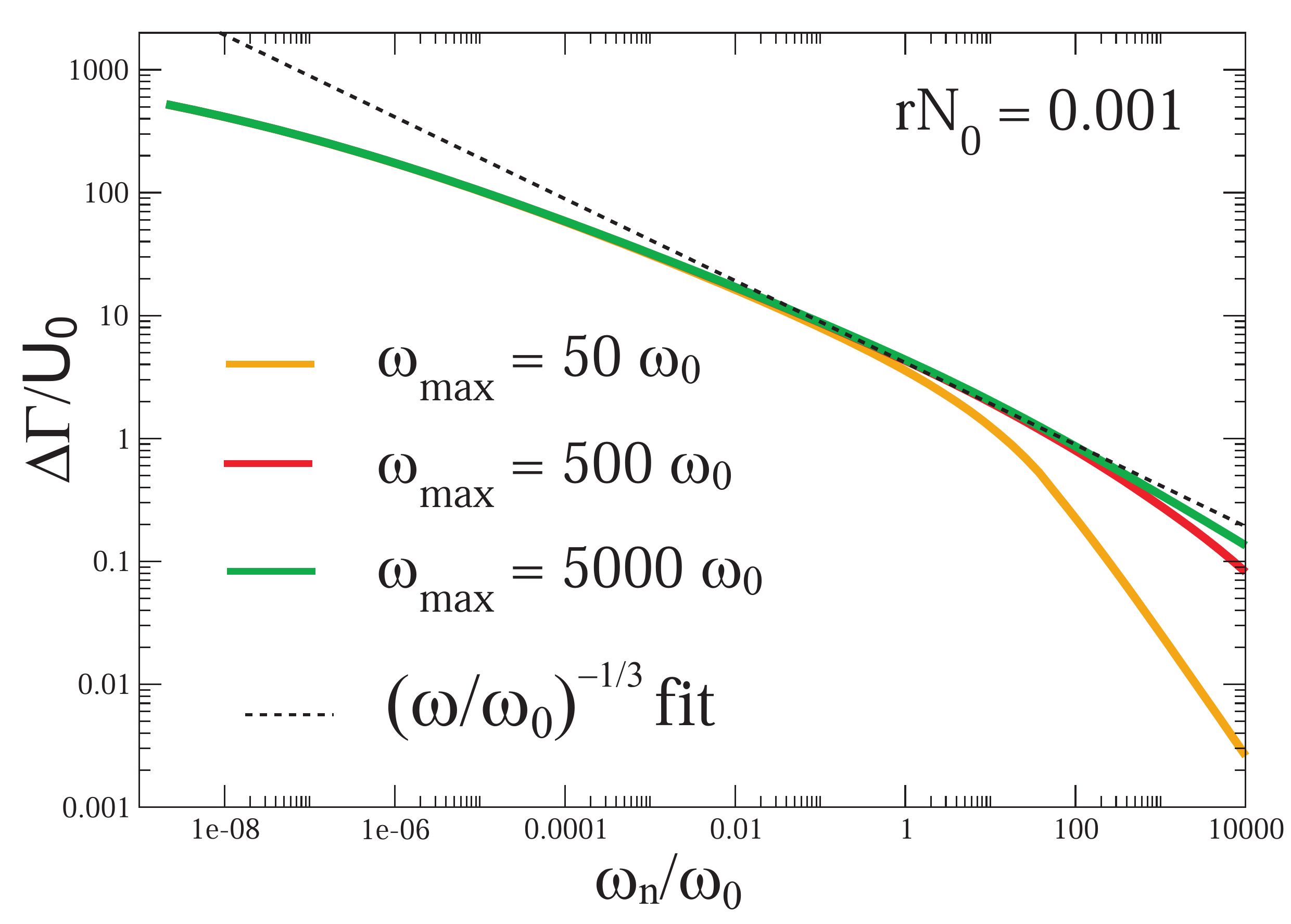}
  \caption{$\dg/U_0$ in the FL case for backward scattering in $D=2$ and several values of the frequency
  cut-off, $\omega_{max}=v_F q_{max}$ together with a fit to $\dg \sim \omega_n^{-1/3}$.}
  \label{fig:2D_bw_hf}
 \end{center}
\end{figure}

In the high-frequency limit, $\omega_n \gg \omega_r$, the leading
frequency dependence is given by $\dg \sim \omega_n^{-1/3}$ [see
Eq.(\ref{eq:dg_2d_bw_hf})]. There are two reasons why this behavior
is not necessarily observed in the numerical results shown in
Fig.~\ref{fig:2D_bw_FL}(c). First, $\dg \sim \omega_n^{-1/3}$ holds
only for the case of an infinitely large fermionic band. In
contrast, for the numerical evaluation of $\dg$, it is necessary to
introduce a finite cut off in the frequency ($\omega_{max}$) and
momentum ($q_{max}$) integration (with $\omega_{max}=v_F q_{max}$)
which establishes a finite fermionic bandwidth (we set
$\omega_{max}=5000 \omega_0$ for the results shown in
Fig.\ref{fig:2D_bw_FL}). Once $\omega_n$ exceeds $\omega_{max}$,
$\dg$ decrease more rapidly than $\omega_n^{-1/3}$. In order to
further demonstrate the dependence of $\dg$ on the $\omega_{max}$ we
present in Fig.~\ref{fig:2D_bw_hf} the frequency dependence of $\dg$
for $r N_0=0.001$ and several values of $\omega_{max}$.
As $\omega_{max}$ increases, $\dg$ decreases slower and extends its
$\omega_n^{-1/3}$-behavior towards higher frequencies, as expected
from the above discussion. Second, in order to observe $\dg \sim
\omega_n^{-1/3}$, it is necessary for the subleading
$\omega_n^{-1/2}$-term [see Eq.(\ref{eq:dg_2d_bw_hf})] to be
negligible in comparison to the leading $\omega_n^{-1/3}$-term. We
find that the frequency, above which the subleading term is
negligible, can be much larger than $\omega_{max}$. To quantify this
result, consider the frequency $\omega_\beta=(3/8z)^3 \beta^6
\omega_0$ such that for $\omega_n>\omega_\beta$, the ratio of the
subleading to leading frequency term in $\dg$ is smaller than
$\beta$. Since, $z \sim 1/r$, one immediately finds that for values
of $rN_0$ of order $O(1)$, $\omega_\beta$ exceeds $\omega_{max}$
even for small values of $\beta$. This explains why for values of
the curvature such as $rN_0=1$ or $rN_0=0.1$ [see
Fig.\ref{fig:2D_bw_FL}(c)], $\dg \sim \omega_n^{-1/3}$ is only
observed as a crossover behavior between the low-frequency
logarithmic divergence and the more rapid decrease at high
frequencies. As $r N_0$ decreases, the frequency range over which
$\dg \sim \omega_n^{-1/3}$ is observed increase because both
$\omega_r$ and $\omega_\beta$ decrease, while $\omega_{max}$ remains
unchanged.

For the NFL case, the derivation of $\dg$ is presented in Appendix
\ref{A1}. For $\omega_n < \omega_0$ one obtains [see
Eq.(\ref{eq:App2DNFLbw})] \beq
 \dg = G(r) \ln\left( \frac{\omega_0}{\omega_n}
 \right).
 \label{eq:dg_2d_bw_nfl}
\enq
where the full form of $G(r)$ is given in Eq.(\ref{eq:G}) (the
logarithmic frequency dependence is similar to the one obtained in
the context of a $U(1)$ gauge theory \cite{Alt95}). In the NFL case,
there exists no crossover scale (such as $\omega_r$ in the FL case)
to an algebraic dependence of $\dg$ below $\omega_0$. However, since
the upper bound for NFL behavior in the fermionic propagator is set
by $\omega_0$, one finds that for $\omega_n \gg \omega_0$, $\dg$
exhibits the same frequency dependence as was obtained for the FL
case [see Eq.(\ref{eq:dg_2d_bw_hf})]. Hence, in the NFL case, the
crossover scale from a logarithmic to an algebraic dependence of
$\dg$ is set by the larger one of $\omega_0$ and $\omega_r$.

\begin{figure}[!h]
 \begin{center}
  \includegraphics*[width=8.5cm]{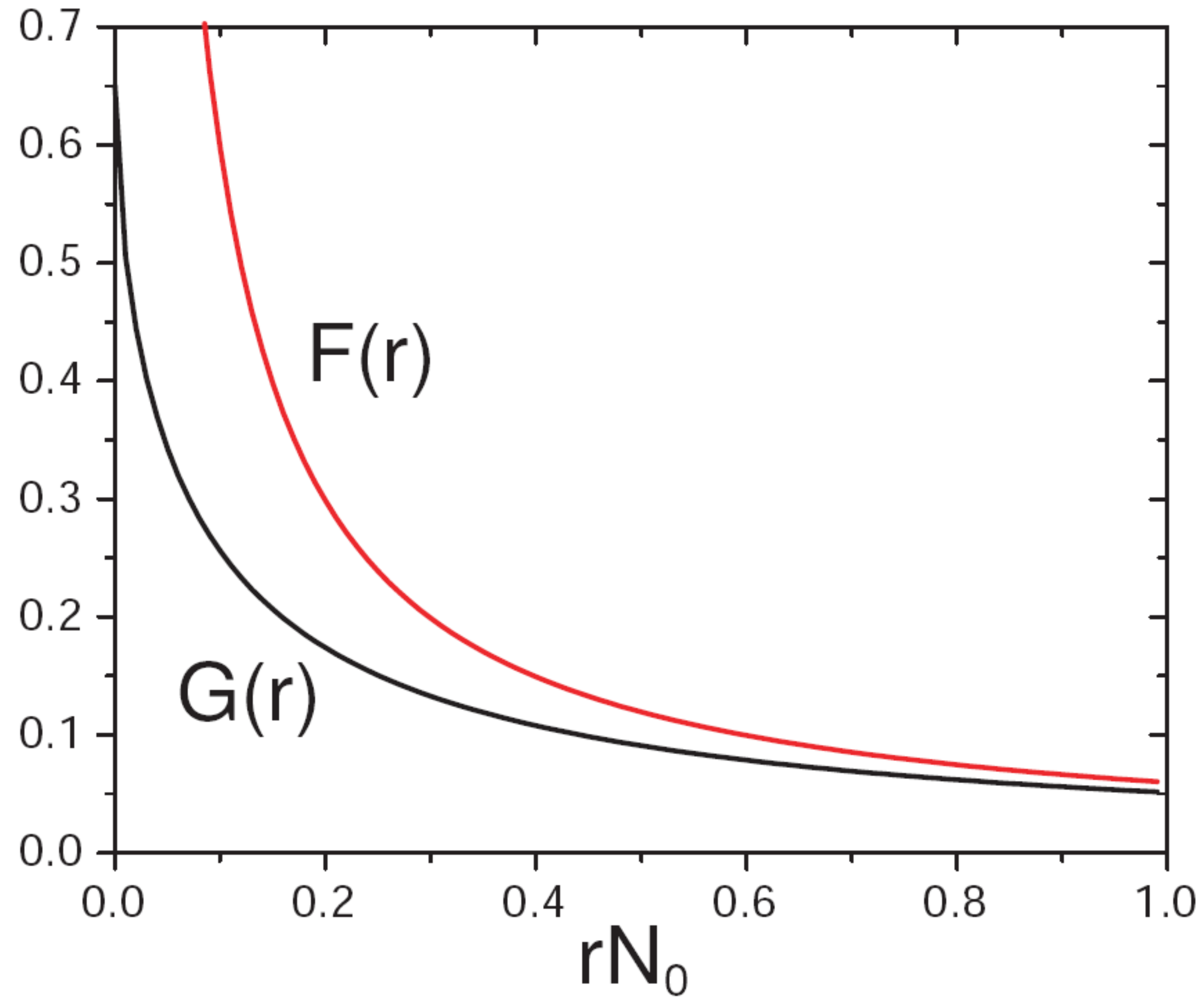}
  \caption{$G(r)$ and $F(r)$ as a function of $r N_0$.}
  \label{fig:G}
 \end{center}
\end{figure}
In Fig.~\ref{fig:G} we present the prefactors of the logarithmic
frequency dependence for the NFL and FL cases, $G(r)$ and
$F(r)=3/(16 \pi r N_0)$, respectively, as a function of curvature.
In both cases, the overall scale of the logarithmic divergence,
rapidly decreases with increasing $r$. Since for all $r$,
$F(r)>G(r)$, we conclude that the inclusion of the fermionic
self-energy [see Eq.(\ref{eq:sigma2D})] which renders the fermionic
Greens function non-Fermi-liquid like, leads to a suppression of the
overall scale of the vertex correction, $\dg$. Moreover, in the
limit of vanishing curvature, $r \rightarrow 0$, one has $G(r)
\rightarrow 2/3$. Hence, the prefactor of the log-divergence is
finite, in contrast to the FL case where $F(r)$ diverges as $\sim
r^{-1}$. In the opposite limit, $r \rightarrow \infty$, one finds to
leading order $G(r) \approx 3/(16 \pi r N_0)$, and hence $G(r)$
approaches $F(r)$ asymptotically.

\begin{figure}[!h]
 \begin{center}
  \includegraphics*[width=8.5cm]{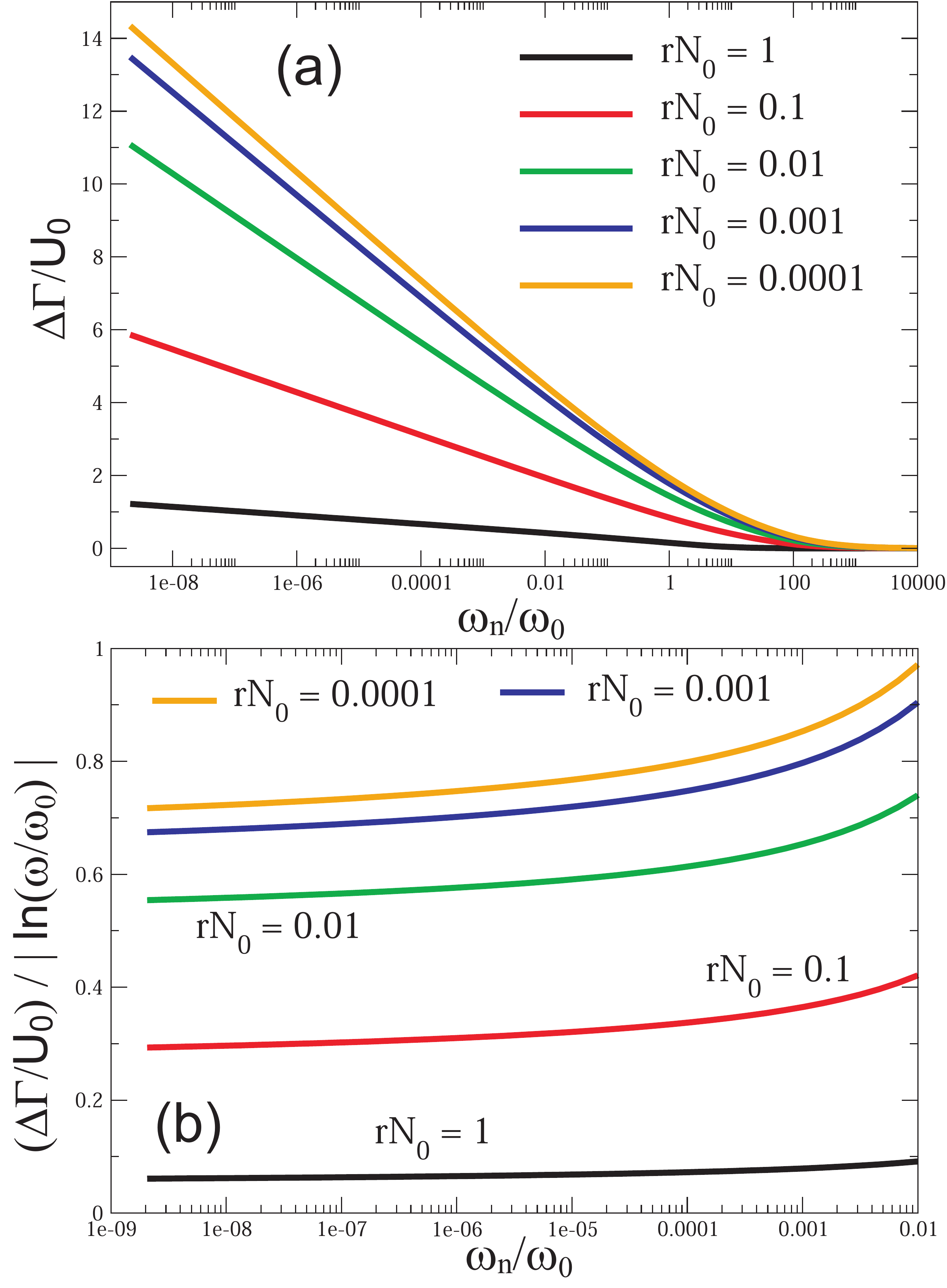}
  \caption{Backward scattering in $D=2$ for the NFL case: (a) $\Delta \Gamma/U_0$,
  (b) $(\Delta \Gamma/U_0)/|\ln \left(\omega_n/\omega_0\right)|$.}
  \label{fig:2D_bw_NFL}
 \end{center}
\end{figure}

In Fig.~\ref{fig:2D_bw_NFL}(a), we present the frequency dependence
of $\Delta \Gamma$ for backward scattering ($\varphi=\pi$) as
obtained numerically from Eq.(\ref{eq:dg_2D}) for the NFL case and
different values of curvature $r$.
In agreement with our analytical results in
Eq.(\ref{eq:dg_2d_bw_nfl}) we find that $\dg$ diverges
logarithmically for $\omega_n \rightarrow 0$. This conclusion is
supported by Fig.~\ref{fig:2D_bw_NFL}(b), where we again plot the
ratio $(\dg/U_0)/ |\ln (\omega_n/\omega_0)|$ which approaches a
constant value for $\omega_n \rightarrow 0$. Moreover, our numerical
results confirm that (a) in the limit of vanishing curvature, the
prefactor of the logarithmic divergence approaches a finite value,
and (b) that the overall scale of the vertex corrections is
suppressed with increasing curvature. Finally, a comparison of the
results for the FL case in Fig.~\ref{fig:2D_bw_FL}(a) with those of
the NFL case in Fig.~\ref{fig:2D_bw_NFL}(a) support our earlier
conclusion that the NFL nature of the fermionic Greens function
suppresses the overall scale of $\dg$.

\subsection{Angular dependence of $\dg$}

\begin{figure}[!h]
 \begin{center}
  \includegraphics*[width=8cm]{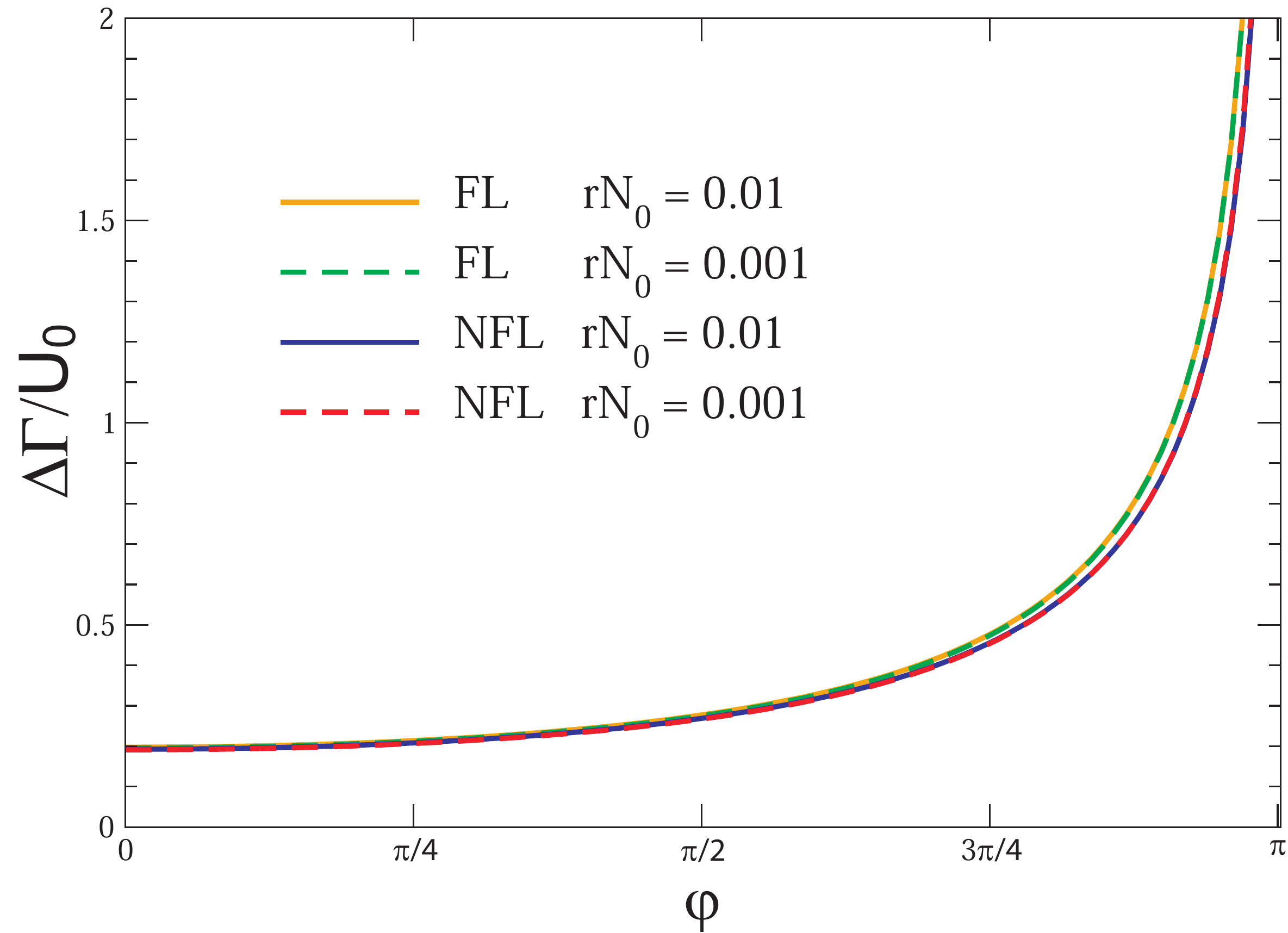}
  \caption{$\dg/U_0$ for $\omega_n=2 \times 10^{-9}\omega_0$ as a function of the
scattering angle, $\varphi$, for the NFL and FL cases in $D=2$ and
several values of $r$.
          }
  \label{fig:2D_phi}
 \end{center}
\end{figure}

In Fig.~\ref{fig:2D_phi}, we present $\dg$ for $\omega_n=2 \times
10^{-9}\omega_0$ as a function of the scattering angle, $\varphi$ in
the FL and NFL cases.
In both cases, $\dg$ increases monotonically with increasing
$\varphi$. Moreover, $\dg$ is practically frequency independent upto
frequencies of order $\omega_0$ over a wide range of scattering
angles (not shown). Only in the immediate vicinity of backward
scattering ($\varphi \approx \pi$) does the frequency dependence of
$\dg$ depend very sensitively on the scattering angle $\varphi$, as
we discuss next.

In order to gain analytical insight into the form of $\dg$ in the
vicinity of backward scattering, we consider the case $r=0$ for
which a full analytical expression of $\dg$ can be obtained (the
derivation is similar to the one discussed in Appendix \ref{A1}). We
begin by discussing the FL case, where in the low frequency limit,
$\omega_n \rightarrow 0$, one obtains
\begin{equation}
 \frac{\dg}{U_0} = \frac{3 \pi}{4} \sqrt{\frac{\alpha}{\phi}} \ ,
 \label{eq:dg_2d_phi_fl_lf}
 \end{equation}
 with $\phi=\pi-\varphi$ such that $\phi=0$ corresponds to backward
 scattering. In the high-frequency limit, $\omega_n \gg \omega^{FL}_\phi$, we have
 \begin{equation}
 \frac{\dg}{U_0} = 2 \left( \frac{\omega_n}{\omega_0}
 \right)^{-1/3} \ ,
 \label{eq:dg_2d_phi fl_hf}
\end{equation}
with the crossover scale being set by
\begin{equation}
\omega_{\phi}^{FL}=z^{-3/2} \omega_0 = \left( \frac{2 \phi}{3\alpha}
\right)^{3/2} \omega_0 \ , \label{eq:2D_phi_crossover}
\end{equation}
which decreases with $\phi$. We thus find that a non-zero angle
$\phi$ eliminates the low-frequency logarithmic divergence of $\dg$
[see Eq.(\ref{eq:dg_2d_bw_lf})] and leads to a constant value of
$\dg$ for $\omega_n \rightarrow 0$. In contrast, the high-frequency
form of $\dg$ remains unchanged from that for backward scattering
($\phi=0$) shown in Eq.(\ref{eq:dg_2d_bw_hf}).

%
\begin{figure}[!h]
 \begin{center}
  \includegraphics*[width=8cm]{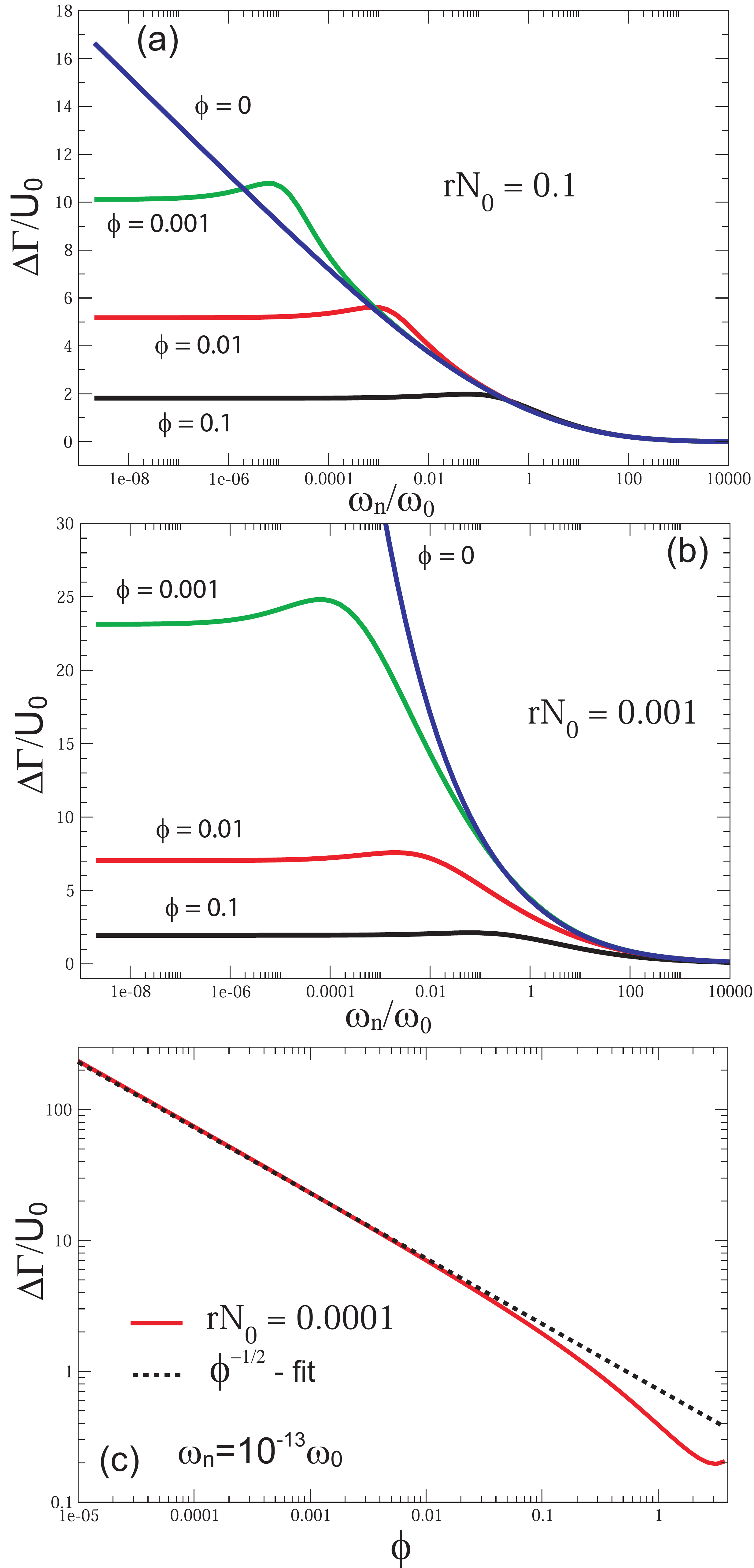}
  \caption{$\dg/U_0$ in $D=2$ as a function of $\omega_n$ for the FL case and several values of $\phi$: (a) $rN_0=0.1$, and (b) $rN_0=0.001$.
  (c) $\dg/U_0$ as a function of $\phi$ for $rN_0=0.0001$ and $\omega_n=10^{-13} \omega_0$.}
  \label{fig:2D_fl_phi}
 \end{center}
\end{figure}
In Figs.~\ref{fig:2D_fl_phi}(a) and (b) we present the frequency
dependence of $\dg$ for the FL case, as obtained from the numerical
evaluation of Eq.(\ref{eq:dg_2D}), for several values of $\phi$ and
curvature $r$. For comparison, we have also included the results for
backward scattering, corresponding to $\phi=0$. In agreement with
the analytical results presented above, we find that the logarithmic
divergence is eliminated by a non-zero $\phi$, and that $\dg$
approaches a constant value for $\omega_n \rightarrow 0$. In
addition, for frequencies above some crossover scale, $\dg$ for
$\phi \not = 0$ exhibits the same frequency dependence as that for
$\phi = 0$, as expected from Eq.(\ref{eq:dg_2d_phi fl_hf}). It is,
in general, difficult to estimate a quantitative value for the
crossover scale from the numerical data presented in
Figs.~\ref{fig:2D_fl_phi}(a) and (b), and to compare them directly
with the analytically obtained result. However, taking, for example,
the maxima in $\dg$ as a measure of the crossover scale, we find
that they scale as $\sim \phi^{3/2}$, in agreement with the
analytical result given in Eq.(\ref{eq:2D_phi_crossover}). Finally,
in Fig.~\ref{fig:2D_fl_phi}(c) we present $\dg$ for vanishing
frequency ($\omega_n = 10^{-13} \omega_0$) and small $rN_0=0.0001$,
as a function of deviation from backward scattering, $\phi$. We find
that $\dg$ scales as $1/\sqrt{\phi}$ [see dotted line in
Fig.~\ref{fig:2D_fl_phi}(c)], in agreement wit the analytical result
presented in Eq.(\ref{eq:dg_2d_phi_fl_lf}).

We next turn to the NFL case, where one obtains for $z \gg 1$ and in
the limit $\omega_n \rightarrow 0$
\begin{equation}
\frac{\dg}{U_0} = 2 \ln \left( \frac{3 \alpha}{2 \phi} \right) ,
\label{eq:dg_2d_nfl_phi_lf}
\end{equation}
in qualitative agreement with the result obtained by Kim and
Millis~\cite{Kim:2003}. In the intermediate frequency regime,
$\omega^{NFL}_\phi \ll \omega_n \ll \omega_0$, one has to leading
order
\begin{equation}
\frac{\dg}{U_0} = \frac{2}{3} \ln \left( \frac{\omega_n}{\omega_0}
\right) \ . \label{eq:dg_2d_nfl_phi_hf}
\end{equation}
Here, the crossover scale is set by
\begin{equation} \omega^{NFL}_\phi=\left(\frac{2 \phi}{3\alpha}
\right)^3 \omega_0 \ .
\end{equation}
For $\omega_n \gg \omega_0$, $\dg$ again takes the FL form given in
Eq.(\ref{eq:dg_2d_bw_hf}). Similar to the FL case, we find that a
non-zero $\phi$ eliminates the logarithmic divergence for $\omega_n
\rightarrow 0$. In addition, there exists an intermediate frequency
range ($\omega^{NFL}_\phi \ll \omega_n \ll \omega_0$) in which $\dg$
exhibits a logarithmic frequency dependence, in contrast to the FL
case. Note that it was argued in Ref.~\cite{Kim:2003} that the
logarithmic dependence of $\dg$ on $\phi$ is responsible for the
anomalies observed in the residual resistivity of ${\rm
Sr_3Ru_2O_7}$ at the metamagnetic transition.

\begin{figure}[!h]
 \begin{center}
  \includegraphics*[width=8cm]{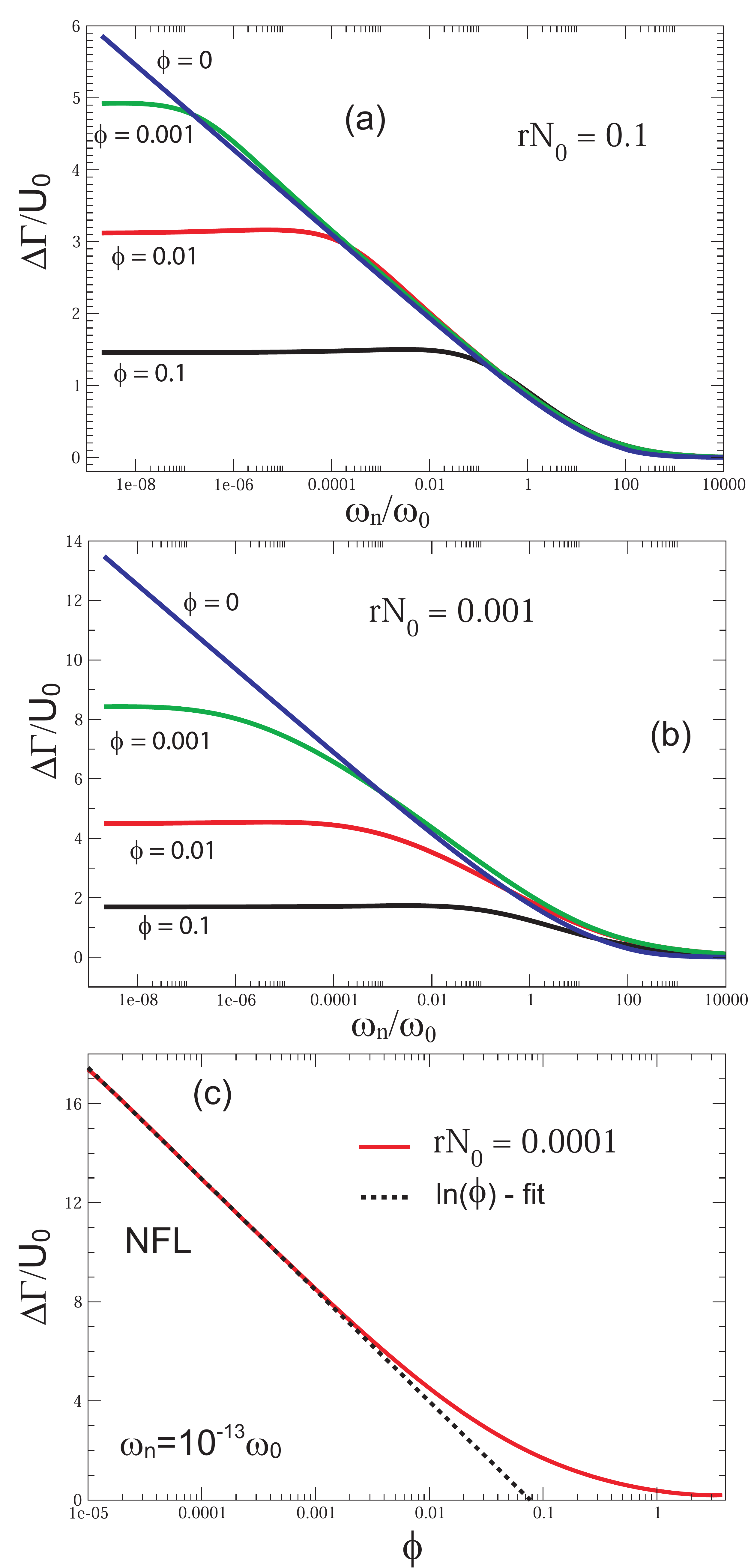}
  \caption{$\dg/U_0$ in $D=2$ as a function of $\omega_n$ for the NFL case and several values of $\phi$: (a) $rN_0=0.1$, and (b)
  $rN_0=0.001$. (c) $\dg/U_0$ as a function of $\phi$ for $rN_0=0.0001$ and $\omega_n=10^{-13} \omega_0$.
          }
  \label{fig:2D_nfl_phi}
 \end{center}
\end{figure}

In Figs.~\ref{fig:2D_nfl_phi}(a) and(b) we present the frequency
dependence of $\dg$ for the NFL case obtained from the numerical
evaluation of Eq.(\ref{eq:dg_2D}), for several values of $\phi$ and
$r$.
In agreement with our analytical results we find that for non-zero
$\phi$, $\dg$ saturates to a finite value in the limit $\omega_n
\rightarrow 0$, and that the crossover scale (below which $\dg$
becomes approximately constant) increases with $\phi$. Above the
crossover scale, the form of $\dg$ for backward scattering
($\phi=0$) and away from backward scattering ($\phi \not =0$) are
practically identical. For small values of $r N_0$, and in the limit
$\phi \rightarrow 0$, we find that $\dg$ scales as $\sim \ln{\phi}$
[see dotted line in Fig.~\ref{fig:2D_nfl_phi}(c)], in agreement wit
the analytical result presented in Eq.(\ref{eq:dg_2d_nfl_phi_lf}).
Moreover, a comparison of Figs.~\ref{fig:2D_fl_phi} and
\ref{fig:2D_nfl_phi} confirms two important analytical results.
First, the limiting value of $\dg$ for $\omega_n \rightarrow 0$ is
smaller in the NFL than in the FL case [see
Eqs.(\ref{eq:dg_2d_phi_fl_lf}) and (\ref{eq:dg_2d_nfl_phi_lf})].
Second, for a given $\phi$, the crossover scale is smaller for the
NFL case than it is for the FL case, i.e., $\omega^{NFL}_\phi \ll
\omega^{FL}_\phi$. By comparing the result for $\phi=0.001$ in
Figs.~\ref{fig:2D_fl_phi}(a) and (b) [and similarly, in
Figs.~\ref{fig:2D_nfl_phi}(a) and (b)] we note that in both cases
the crossover scale moves to lower frequencies with increasing $r$.

%
\section{Vertex Corrections in $D=3$}
\label{sec:vc_3D}
Since the higher dimensionality leads to a weaker deviation of the
fermionic propagator from Fermi-liquid behavior in $D=3$ than in
$D=2$ (see Sec.\ref{sec:self_energy}), we expect a concurrent weaker
renormalization of the scattering amplitude in $D=3$ as well. For
the numerical results presented below, we set $\gamma =0.1$,
$I_s=3$, and give $r N_0$ in units of $[\frac{\pi}{k_F}
\left(\frac{\Lambda}{k_F} \right)^3]^{-1}$, while $q_{max}$ and
$\omega_{max}$ are given in units of  $q_0 \gamma e^{1/\gamma}$ and
$\omega_0 \gamma e^{1/\gamma}$, respectively.
%
%
\subsection{Forward scattering}
\begin{figure}[h]
 \begin{center}
  \includegraphics*[width=8cm]{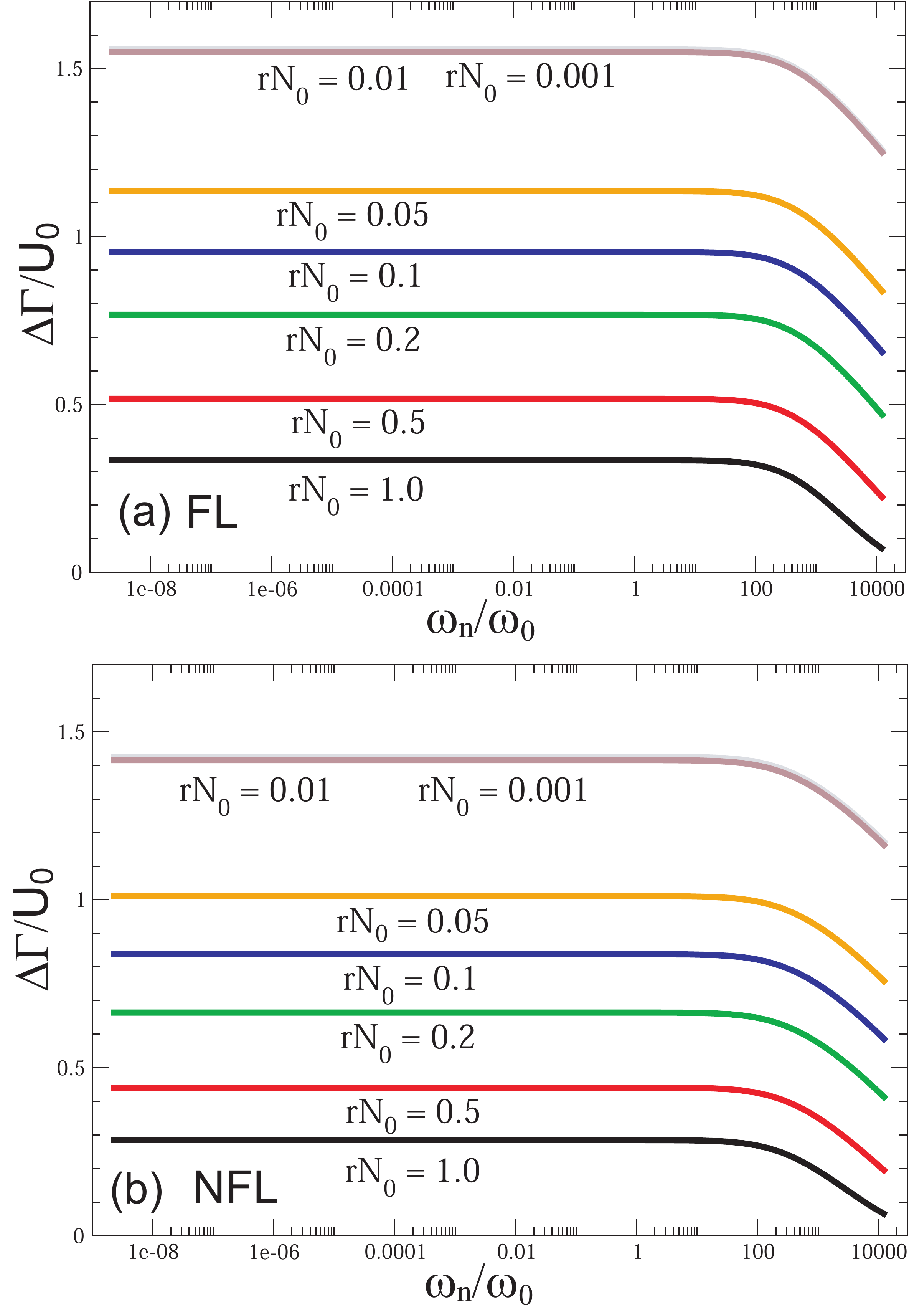}
  \caption{
           Frequency dependence of $\dg/U_0$ in $D=3$ for forward scattering and several values of
           $rN_0$: (a) the FL case, and (b) the NFL case (with
           $\omega_{max}=q_{max}=100$).
          }
  \label{fig:3D_fw}
 \end{center}
\end{figure}

\begin{figure}[h]
 \begin{center}
  \includegraphics*[width=8cm]{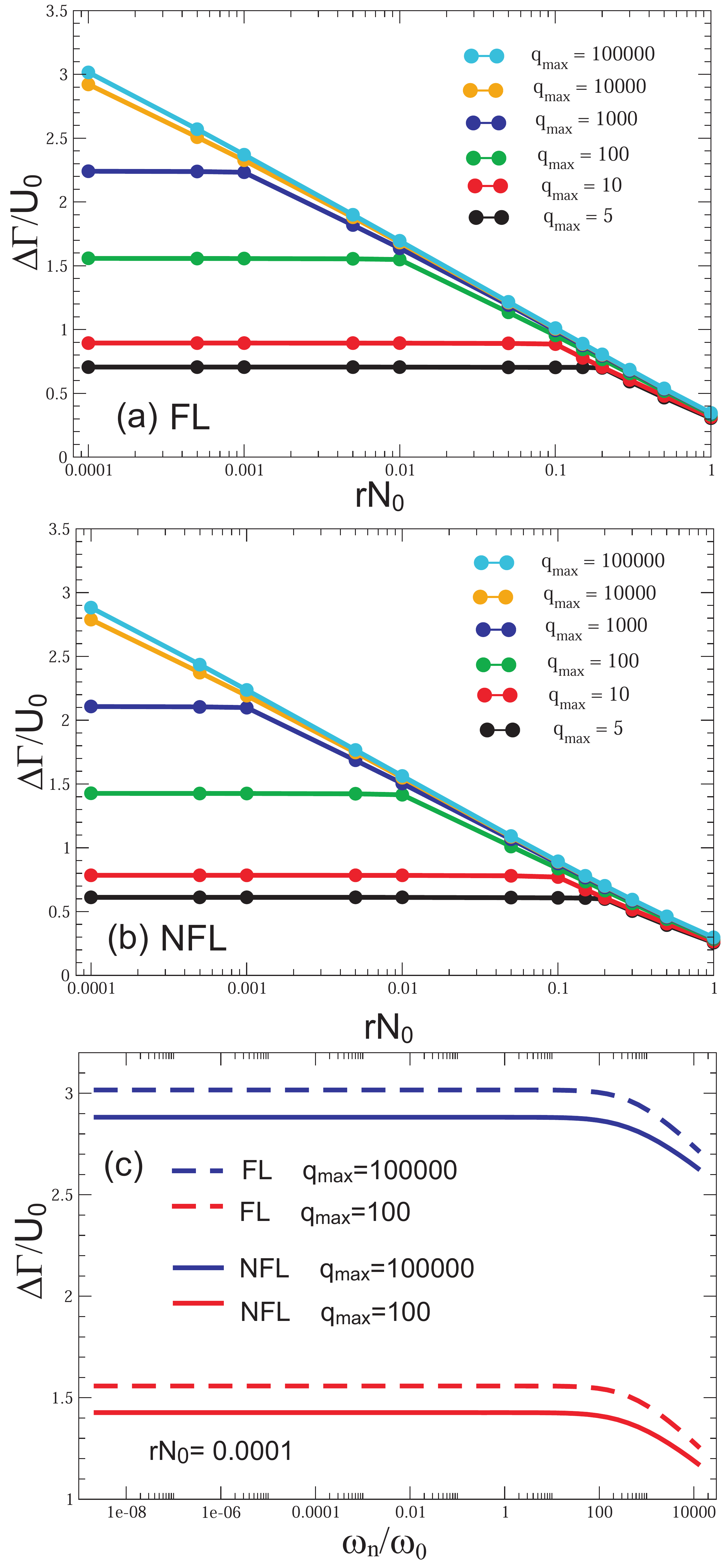}
  \caption{
           $\dg/U_0$ for $\omega_n
=2 \times 10^{-9}$ and forward scattering in $D=3$ as a function of
$rN_0$: (a) the FL case, and (b) the NFL case. (c) Frequency
dependence of $\dg/U_0$ for $rN_0=0.0001$ and several values of
$q_{max}$.}
  \label{fig:3D_fw_r}
 \end{center}
\end{figure}

The frequency dependence of $\dg$ for forward scattering, obtained
from the numerical evaluation of Eq.(\ref{eq:dg_3D}) is shown in
Figs.~\ref{fig:3D_fw}(a) and (b), for the FL and NFL cases,
respectively (similar to our results in $D=2$, we note that the
non-zero contribution to $\dg$ for forward scattering obtained from
the numerical evaluation arises from the branch cut in $\chi$ and
the finite cut-off in the momentum integration).
In both cases, $\dg$ is practically frequency independent for
frequencies below $\omega_0$, and hence approaches a constant value
in the limit $\omega_n \rightarrow 0$. We note, however, that the
overall scale of $\dg$ increases with decreasing $r$, upto a value,
$r_c$, below which $\dg$ becomes independent of $r$ [for example,
$\dg$ for $rN_0=0.01$ and $0.001$ are practically indistinguishable
in Figs.~\ref{fig:3D_fw}(a) and (b).] We find from an analytical
analysis of Eq.(\ref{eq:dg_3D}) that $r_c N_0=1/q_{max}$ where
$q_{max}$ is the upper cut-off in the momentum integration of
Eq.(\ref{eq:dg_3D}). This result is in agreement with our numerical
analysis which is summarized in Fig.~\ref{fig:3D_fw_r} where we
present $\dg$ at $\omega_n=2 \times 10^{-9} \omega_0$ as a function
of $rN_0$ for several $q_{max}$.
As follows from Figs.~\ref{fig:3D_fw_r}(a) and (b), one has $\dg
\sim \ln(r)$ down to $r_c$ below which $\dg$ becomes independent of
$r$. In order to investigate whether the functional form of the
frequency dependence of $\dg$ changes between $r<r_c$ and $r > r_c$,
we present in Fig.~\ref{fig:3D_fw_r}(c) $\dg$ as a function of
frequency for $rN_0=0.0001$ and different values of $q_{max}$.  For
$q_{max}=100$, one has $r<r_c$, while $r>r_c$ for $q_{max}=10^5$. We
find not only that the functional form of $\dg$ is practically
independent of whether $r$ is smaller or larger than $r_c$, but also
that the crossover frequency below which $\dg$ becomes frequency
independent does not change with $r$ or $q_{max}$.  This result is
in apparent contradiction to the findings of Miyake {\it et al.}
\cite{Miyake:2001} who argued that in the FL case, $\dg$ diverges
logarithmically for $\omega_n \rightarrow 0$. At present, the origin
for this discrepancy is unclear. Our result, however, is consistent
with the finding that in $D=2$, $\dg$ also approaches a finite value
for $\omega_n \rightarrow 0$, since, in general, one would expect
that with increasing dimensionality of the system, fluctuation
corrections become weaker, and that as a result, the functional
dependence of $\dg$ on $\omega_n$ should not become stronger in the
low frequency limit.

%
\subsection{Backward scattering}

In order to gain analytical insight into the functional form of
$\dg$ for backward scattering in $D=3$, we again start from
Eq.(\ref{eq:an}). In the FL case and in the limit $\omega_n
\rightarrow 0$, one finds to leading order
\begin{equation}
\frac{\dg}{U_0} = 9 \pi^2 \gamma \left[ F\left( x
\right)-\frac{\sqrt{2}}{3 \rho^{1/2}} \sqrt{
\frac{\omega_n}{\omega_0}} + \frac{2}{3 \sqrt{3} \rho^{2/3}}
\left(\frac{\omega_n}{\omega_0} \right)^{2/3} \right]
 \label{eq:3d_fl_scaling}
\end{equation}
where $\rho=\lambda^2 r^3/(\langle v_F \rangle ^2 \omega_0 )$,
$x=\left( \rho^{-1} \omega_{max}/\omega_0 \right)^{1/3}$,
$\omega_{max}$ is the upper cut-off in frequency, and $F(x)$ is a
universal function of the upper cut-off, which scales as $F(x)
\approx \ln^2(x)/(2\pi)$ for $x \to\infty$, i.e. $r\to 0$. In the
NFL case we obtain to leading order
\begin{equation}
 \frac{\dg}{U_0} = \frac{3}{2} \left[ F\left(x \right)
 -\frac{\sqrt{2}\pi}{\rho^{1/2}} \sqrt{\left(
 \frac{\omega_n}{\omega_0} \right)
 \ln{\left(\frac{\omega_0}{\omega_n}\right)}} \right]
 \label{eq:3d_nfl_scaling}
\end{equation}
with $F(x)$ given above. Thus, the zero frequency limit of $\dg$ for
both the FL and NFL cases is finite, though it diverges with
vanishing curvature $r$ as $\dg \sim \ln^2(r)$.

\begin{figure}[!h]
 \begin{center}
  \includegraphics*[width=8cm]{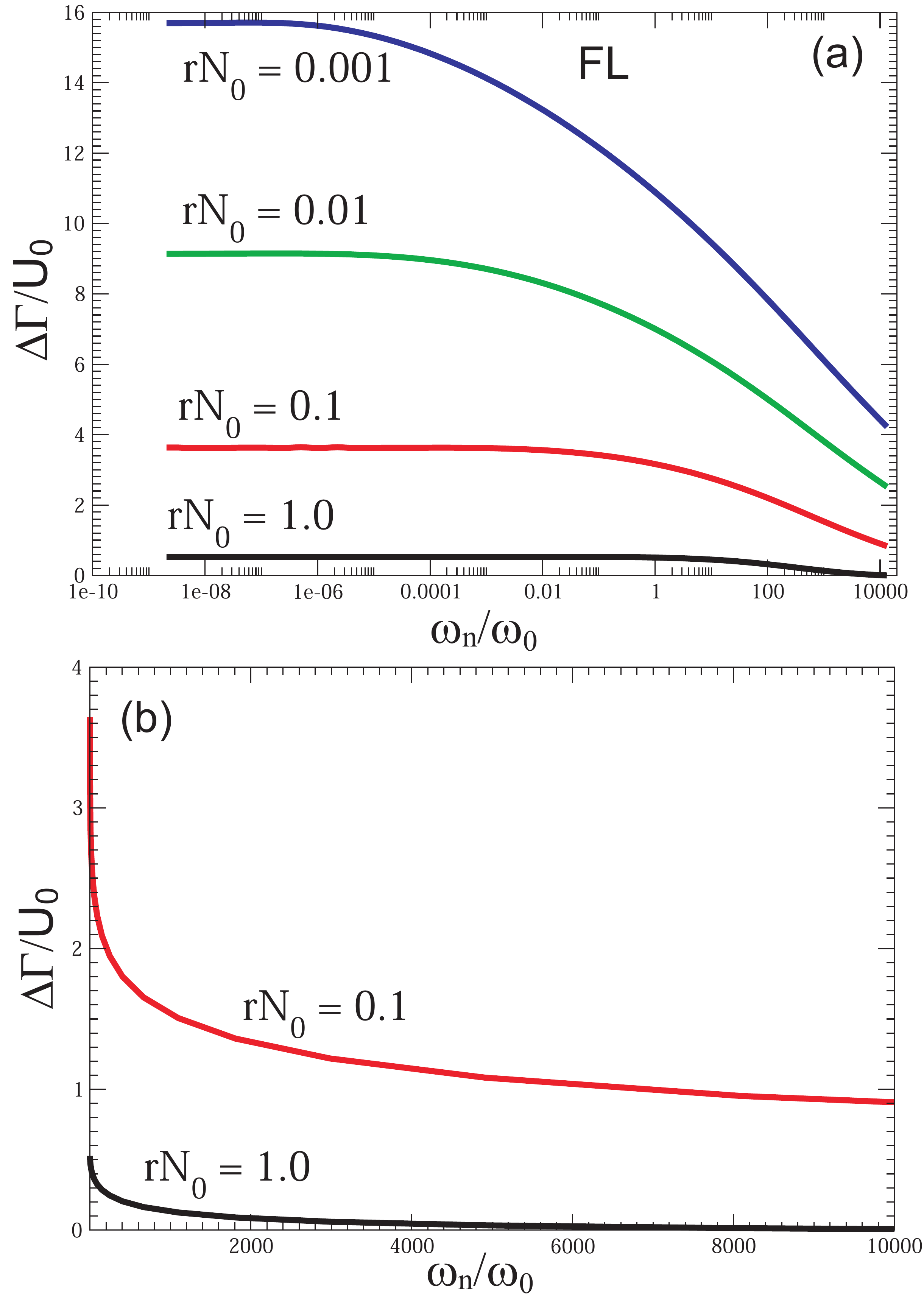}
  \caption{
           $\dg/U_0$ in $D=3$ for backward scattering in the FL case for different values of $rN_0$: (a) linear-log plot,
           (b) same results as in (a) but on a linear-linear plot.
           We set $\omega_{max}=q_{max}=10$.
          }
  \label{fig:3D_bw_FL}
 \end{center}
\end{figure}

\begin{figure}[!h]
 \begin{center}
  \includegraphics*[width=8cm]{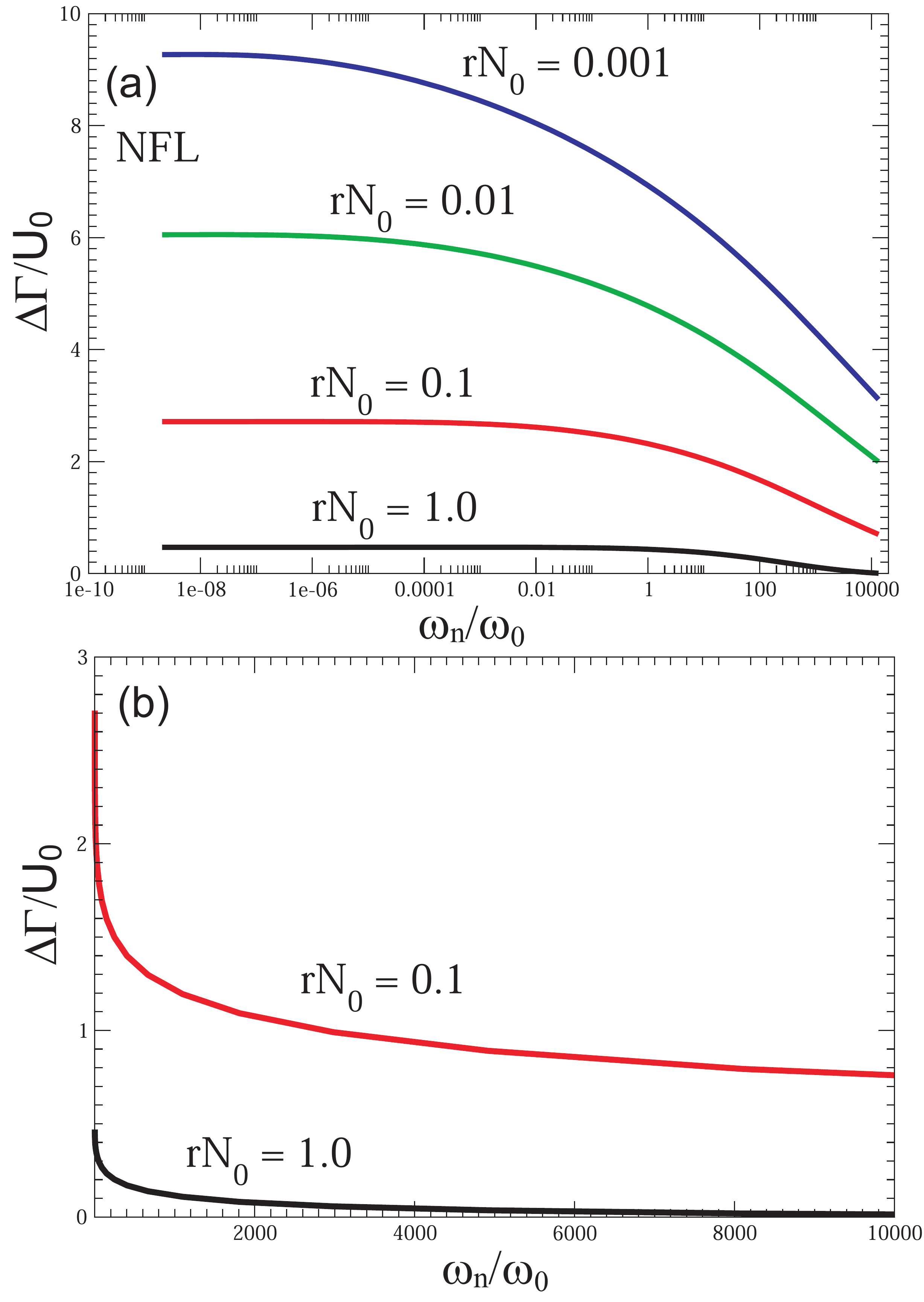}
  \caption{
            $\dg/U_0$ in $D=3$ for backward scattering in the NFL case for different values of $rN_0$: (a) linear-log plot,
           (b) same results as in (a) but on a linear-linear plot. We set $\omega_{max}=q_{max}=10$.
          }
  \label{fig:3D_bw_NFL}
 \end{center}
\end{figure}

In Figs.~\ref{fig:3D_bw_FL} and \ref{fig:3D_bw_NFL} we present $\dg$
in the FL and NFL cases, respectively, for backward scattering
($\varphi=\pi$), as obtained from the numerical evaluation of
Eq.(\ref{eq:dg_3D}).
In agreement with the analytical results, we find that in both
cases, $\dg$ approaches a constant value as $\omega_n \rightarrow
0$, in contrast to the logarithmic divergence found in $D=2$.
Moreover, our numerical results show that $\dg$ at the lowest
frequency scales as $\sim \ln^2(r)$ (in contrast to $\sim \ln(r)$
for forward scattering), and that the leading frequency correction
in the FL case is given by $\sim \sqrt{\omega_n}$, in agreement with
our analytical findings.

\subsection{Angular dependence of $\dg$}

\begin{figure}[!h]
 \begin{center}
  \includegraphics*[width=8cm]{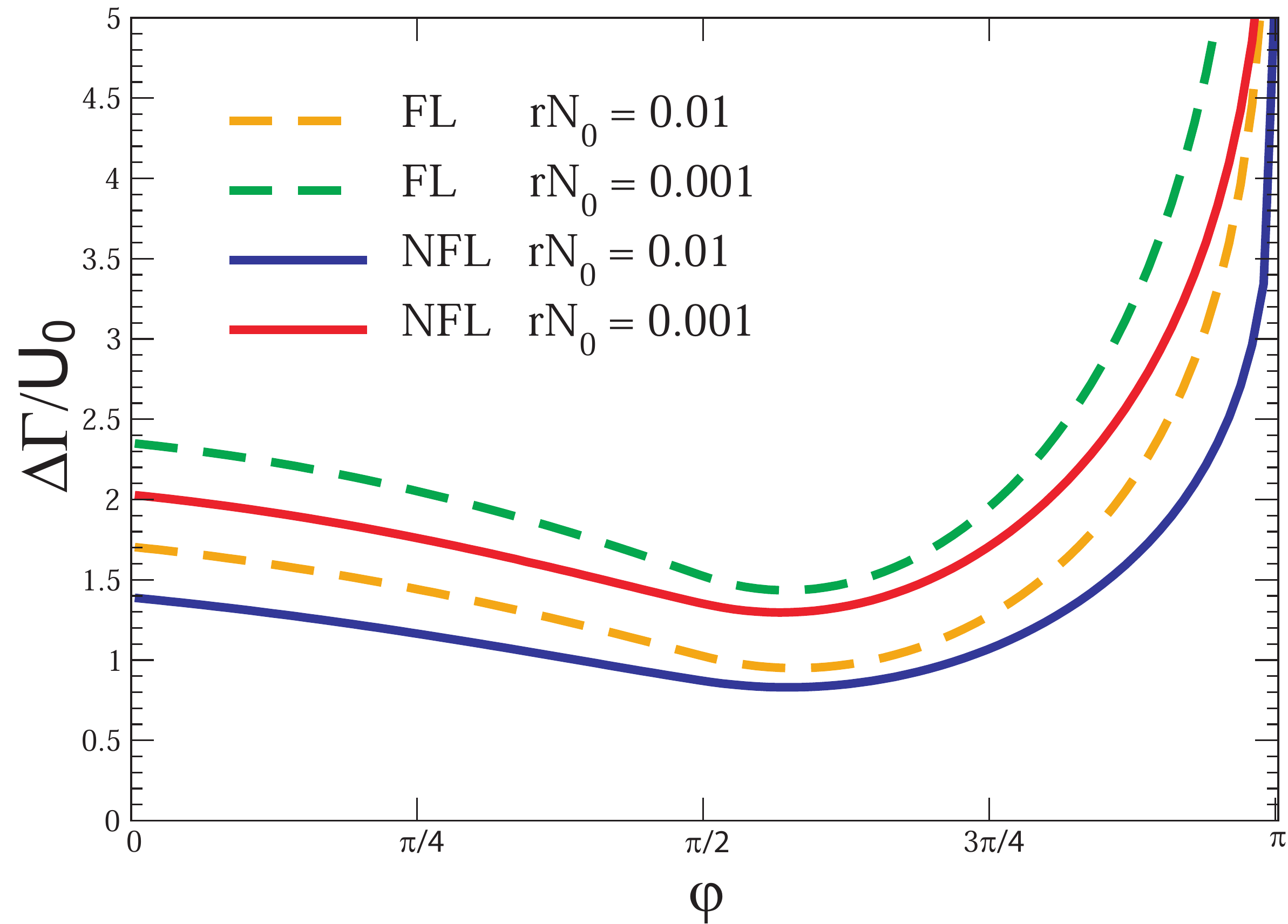}
  \caption{$\dg/U_0$ for $\omega_n=2 \times 10^{-9} \omega_0$ as a function of
scattering angle $\varphi$ in $D=3$ for the NFL and FL cases and
several values of $rN_0$. We set $\omega_{max}=q_{max}=10^4$.
          }
  \label{fig:3D_phi}
 \end{center}
\end{figure}

In Fig.~\ref{fig:3D_phi}, we present $\dg$ for $\omega_n=2 \times
10^{-9} \omega_0$ as a function of scattering angle $\varphi$ in the
FL and NFL cases for several values of $r$.
In both cases, $\dg$ first decreases with increasing $\varphi$,
exhibiting a minimum around $\varphi \approx 0.578 \pi$, and then
increases sharply as $\varphi$ approaches $\pi$. Similar to the case
in $D=2$, we find that  $\dg$ is practically frequency independent
upto frequencies of order $\omega_0$ over a wide range of scattering
angles. Only in the immediate vicinity of backward scattering
($\varphi \approx \pi$) does the frequency dependence of $\dg$
depend very sensitively on the scattering angle $\varphi$. A
comparison of Fig.~\ref{fig:2D_phi} with Fig.~\ref{fig:3D_phi} show
that the dependence of $\dg$ on curvature $r$ is significantly
stronger in $D=3$ that it is in $D=3$, in agreement with the above
analytical results.

%
\section{Higher Order Vertex Corrections}
\label{sec:HigherOrder}

We next discuss the form of higher order vertex corrections, such as
the one shown in Fig.~\ref{fig:vcdiag}~(c). It turns out that the
low-frequency form of the higher order vertex corrections is
entirely determined by that of the lowest order vertex correction,
$\Delta \Gamma^{(1)} \equiv \dg$ discussed in the previous sections.
In what follows, we consider the case of vanishing curvature, where
the segments between wavy lines of the higher order (ladder) vertex
corrections decouple, and the infinite series of ladder vertex
corrections can be summed up.

To demonstrate this, we first consider the case of backward
scattering in $D=2$ for the NFL case, where in the low-frequency
limit, we found for the lowest order vertex correction
$\dg^{(1)}=\dg=2/3 \ln(\omega_0/\omega_n)$ [see
Eq.(\ref{eq:dg_2d_bw_nfl}) for $r=0$]. It then immediately follows
that the leading frequency dependence of the $m'th$ order vertex
correction in the limit $\omega_n \rightarrow 0$ is given by
\begin{equation}
\Delta \Gamma^{(m)} = \frac{1}{m!} \left[ \frac{2}{3} \ln\left(
\frac{\omega_0}{\omega_n} \right) \right]^m
\end{equation}
Summing up the entire series of ladder vertex corrections (including
the bare vertex) yields for the renormalized scattering vertex for
backward scattering
\begin{equation}
U=U_0 \left( \frac{\omega_n}{\omega_0} \right)^{-2/3}
\end{equation}
Performing the analytical continuation to real frequencies, one
obtains
\begin{equation}
U=U_0 \left( \frac{\omega}{\omega_0} \right)^{-2/3} e^{\sigma i
\pi/3} \label{eq:full_u_1}
\end{equation}
where the choice of the branch in the complex plane ($\sigma=\pm$)
is determined by requirement that the LDOS, renormalized by the
scattering off the impurity, be positive. For a particle-hole
symmetric band structure, we find that this requirement is met by
using $\sigma=-{\rm sgn}(U_0)$. We thus find an algebraic frequency
divergence of the full vertex in the low frequency limit, similar to
the result obtained by Altshuler {\it et al.}~\cite{Alt95} in the
context of a $U(1)$ gauge theory. For zero frequency, it was shown
by Kim and Millis \cite{Kim:2003} that the same summation of ladder
diagrams in $D=2$ for the NFL case leads to similar algebraic
dependence of $\dg$ on the deviation from backward scattering,
$\phi$.

As a second example for the form of the higher order ladder
diagrams, we consider the case where the lowest order vertex
correction, $\dg^{(1)}$, approaches a constant value in the limit
$\omega_n \rightarrow 0$, i.e., $\dg^{(1)}/U_0=A$. In the same
limit, the $m'th$ order vertex correction is given by
\begin{equation}
\frac{\Delta \Gamma^{(m)}}{U_0} = A^m \ .
\end{equation}
Summing up the entire series of ladder vertex corrections (including
the bare vertex) yields for the renormalized scattering vertex for
backward scattering
\begin{equation}
U=U_0 \frac{1}{1-A}
\label{eq:full_u_2}
\end{equation}
As a result, we find that even for those cases where the lowest
order vertex correction does not diverge in the limit $\omega_n
\rightarrow 0$, the full scattering vertex can be enhanced, or even
diverge, depending on the zero-frequency limit of $\dg^{(1)}$.
Finally, note that the (crossing) vertex correction diagrams of the
type shown in Fig.~\ref{fig:vcdiag}~(d) are expected to not
qualitatively modify the results given in \ceq{eq:full_u_1} and
\ceq{eq:full_u_2}, [\onlinecite{Kim:2003}].

%
\section{Conclusions}
\label{sec:conclusion}
\begin{table*}
 \caption{Summary of the leading frequency dependence of $\dg$ for backward
          scattering in the limit $\omega_n \rightarrow 0$.
          All quantities are defined in the main text.
          The function $G$ is given by \ceq{eq:G} and shown in Fig.~\ref{fig:G}.
      As stated in the text, after \ceq{eq:3d_fl_scaling}, $F(x)$ is a universal function,
      which scales as $F(x) \approx \ln^2(x)/(2\pi)$ for $x \to\infty$, where
      $x=r^{-1}(v_F^2\Omega_c/\lambda^2)^{1/3}$.
      For forward scattering we find that
          both in $D=2$ and $3$ the lowest order vertex corrections do not diverge as $\omega_n\to 0$.}
 \label{tab:1}
  \begin{center}
    \begin{tabular}{ | p{1.0cm} || p{7.6cm} | p{8.6cm} |}
    \hline
        & \begin{center}{\bf Fermi Liquid}\end{center} & \begin{center}{\bf Non-Fermi liquid}\end{center} \\
    \hline
    \hline
     \begin{center}{\bf $D=2$}\end{center}
     & $$ \frac{\dg}{U_0} = \frac{3}{16\pi}\frac{1}{rN_0}\ln\left(\frac{\omega_n}{\omega_0}\right) $$
     & $$ \frac{\dg}{U_0} \approx G\left(r\right)\ln\left(\frac{\omega_0}{\omega_n}\right) $$
     \\
    \hline
     \begin{center}{\bf $D=3$}\end{center}
     & $$ \frac{\dg}{U_0} =  9 \pi^2 \gamma \left[ F\left( x
\right)-\frac{\sqrt{2}}{3 \rho^{1/2}} \sqrt{
\frac{\omega_n}{\omega_0}}  \right] $$
     & $$ \frac{\dg}{U_0} = \frac{3}{2} \left[ F\left(x \right)
-\frac{\sqrt{2}\pi}{\rho^{1/2}} \sqrt{\left(
\frac{\omega_n}{\omega_0} \right)
\ln{\left(\frac{\omega_0}{\omega_n}\right)}} \right]$$ \\
    \hline
    \end{tabular}
\end{center}
\end{table*}
In summary, we have studied the renormalization of a non-magnetic
impurity's scattering potential due to the presence of a massless
collective spin mode at a ferromagnetic quantum critical point. For
the case of a single, isolated impurity (corresponding to the limit
of a vanishing impurity density), we computed the lowest order
vertex corrections in two- and three-dimensional systems, for
arbitrary scattering angle, frequency and curvature. We showed that
only for backward scattering, $\varphi = \pi$, in $D=2$ does the
lowest order vertex correction diverge logarithmically in the limit
$\omega_n \rightarrow 0$ (a summary of these results is shown in
Table \ref{tab:1}). A similar result (for the NFL case) was obtained
in the context of a $U(1)$ gauge theory by Altshuler {\it et
al.}~\cite{Alt95}. For $\varphi \not = \pi$ in $D=2$, and for all
$\varphi$ in $D=3$, we find that the vertex corrections both for the
NFL and FL cases approach a finite (albeit possibly large) value in
the low-frequency limit. In particular, in the vicinity of backward
scattering in $D=2$, we find that the logarithmic frequency
divergence in $\dg$ is cut by a non-zero deviation from backward
scattering, $\phi$, with $\dg \sim 1/\sqrt{\phi}$ for the FL case,
and $\dg \sim \ln(\phi)$ in the NFL case. The latter result is in
qualitative agreement with that obtained by Kim and
Millis~\cite{Kim:2003}. Moreover, for the NFL case and forward
scattering in $D=2$, our finding of a finite $\dg$ in the limit
$\omega_n \rightarrow 0$ agrees with that by Rech {\it et al.}
\cite{Rech:2006}. However, our results are in disagreement with
those of Miyake {\it et al.}\cite{Miyake:2001} who for the FL case
reported a logarithmic divergence in frequency of $\dg$ for forward
scattering in $D=3$. The origin of this discrepancy is presently
unclear.

We also showed that the overall scale of vertex corrections is
weaker in the NFL than in the FL case, implying that the NFL nature
of the fermionic degrees of freedom diminish the vertex corrections.
Furthermore, we demonstrated that vertex corrections for backward
scattering are strongly suppressed with increasing curvature of the
fermionic bands; the qualitative nature of this suppression,
however, is different in the NFL and FL cases. Moreover, $\dg$
exhibits in general a stronger dependence on $r$ in $D=3$ than in
$D=2$. In particular, for forward scattering, we find that in $D=2$,
$\dg$ in the low-frequency limit is practically independent of the
curvature, $r$, but that in $D=3$, $\dg$ depends logarithmically on
$r$ down to $r_cN_0=1/q_{max}$ below which it becomes independent of
$r$. We explicitly computed the full angular dependence of the
vertex correction, and showed that they vary only weakly over a
large range of scattering angles and frequencies, but are strongly
enhanced in the vicinity of backward scattering. Finally, we
considered higher order ladder vertex corrections, and showed that
their zero-frequency limit, for $r \to 0$, is solely determined by
that of the lowest order vertex correction. As a result, it is
possible to sum an infinite series of ladder diagrams. We showed
that for backward scattering in $D=2$, this summation changes the
logarithmic frequency dependence of the lowest order vertex
correction into an algebraic frequency dependence of the fully
renormalized scattering vertex. For zero frequency, it was shown by
Kim and Millis \cite{Kim:2003} that a summation of ladder diagrams
in $D=2$ for the NFL case leads to similar algebraic dependence of
$\dg$ on the deviation from backward scattering.

The question naturally arises whether the combined angular and
frequency dependence of the vertex corrections discussed above are
experimentally measurable, for example, via a combination of
frequency dependent and local measurements. For example, by using
several impurities in a well-defined spatial geometry (which could
predominantly probe vertex corrections for certain scattering
angles), scanning tunneling spectroscopy experiments could provide
insight into the combined angular and frequency dependence of vertex
corrections via measurements of the local density of states.
Moreover, since the resistivity of a material is predominantly
determined by backscattering, the frequency dependence of $\dg$ for
backscattering might be experimentally detectable in the optical
conductivity \cite{Belitz:2000}. A theoretical investigation of
these questions is reserved for future studies. We note, however,
that the combination of theoretical and experimental results on the
form of the LDOS and the optical conductivity will provide important
insight into the nature of vertex corrections in particular, and
into the interplay of quantum fluctuations and disorder in general.

%
\section{Acknowledgments}
We thank A. Chubukov, D. Maslov, A. Millis, K. Miyake, and G.
Schwiete for helpful discussions. D.K.M. acknowledges financial
support by the National Science Foundation under Grant No.
DMR-0513415 and the U.S. Department of Energy under Award No.
DE-FG02-05ER46225.


\appendix

\section{Derivation of $\dg$ in $D=2$ for backward scattering}
\label{A1}

The analytical derivation of $\dg$ starts from Eq.(\ref{eq:an}).
Performing the integration over $q_\parallel$ yields
\begin{eqnarray}
 \frac{\dg}{U_0} &=& -\frac{3g^2\chi_0}{(2\pi)^3}\frac{\pi}{v_F}
        \int_{-\infty}^{\infty}d\nu \int_{-\infty}^{\infty}d q_\perp
    \frac{1}{q^2_\perp + \frac{\lambda |\om -\nu|}{v_F |q_\perp|}}  \nonumber \\
        & & \times \frac{1}{F(|\nu|)+irq^2_\perp \sign(\nu) }
 \label{eq:dg_2d_bw3}
\end{eqnarray}
where $F(|\nu|)=|\nu|$ for the FL case and $F(|\nu|)=|\nu|+
\omega_0^{1/3}|\nu|^{2/3}$ for the NFL case. While it is possible to
analytically perform the integrations over $q_\perp$ and $\nu$ in
Eq.(\ref{eq:dg_2d_bw3}), it turns out that a better analytical
understanding of $\dg$ can be obtained by eliminating the frequency
$\omega_n$ from the integrand and reintroducing it as a lower
cut-off in the frequency integration. We explicitly checked
(analytically and numerically) that the resulting leading order
frequency dependence of $\dg$ as discussed above is the same in both
methods. We then obtain (dropping the subscript of $q_\perp$)
\begin{eqnarray}
 \frac{\dg}{U_0} &=& -\frac{12g^2\chi_0}{(2\pi)^3}\frac{\pi}{v_F}
        \int_{\omega_n}^{\infty}d\nu \int_{0}^{\infty}d q
    \frac{q}{q^3 + \frac{\lambda \nu}{v_F}}  \nonumber \\
        & & \times \frac{F^2(|\nu|)}{F^2(|\nu|)+r^2q^4  } \ .
 \label{eq:dg_2d_bw4}
\end{eqnarray}
Rescaling momenta and frequencies as
\begin{equation}
\hat{\omega}=\frac{\omega}{\omega_0}; \, \hat{q}=\frac{q}{q_0}; \,
q_0=\frac{\omega_0}{v_F}; \, {\hat F}=\frac{F}{\omega_0}
\end{equation}
one obtains
\begin{eqnarray}
 \frac{\dg}{U_0} &=& -\frac{12g^2\chi_0}{(2\pi)^3}\frac{\pi}{v_F}
        \int_{\omega_n}^{\Omega_\Lambda} d\nu \int_{0}^{\infty}d {\hat q}
    \frac{{\hat q}}{{\hat q}^3 + \beta^{3} {\hat \nu}}  \nonumber \\
        & & \times \frac{{\hat F}^2(|\nu|)}{{\hat F}^2(|\nu|)+{\hat r}^2{\hat q}^4  }
 \label{eq:dg_2d_bw5}
\end{eqnarray}
where
\begin{equation}
\beta \equiv \frac{4}{3}\alpha^{-1} ; \quad {\hat r}\equiv r
\frac{q^2_0}{ \omega_0}
\end{equation}
We next consider the NFL case. Since the upper frequency cut-off for
NFL behavior is set by $\omega_0$, we can introduce $\omega_0$ as a
high frequency cut-off in the frequency integration, and use the
approximation $F(|\nu|) \approx \omega_0^{1/3}|\nu|^{2/3}$ which is
valid for $\omega_n < \omega_0$ (for $\omega_n>\omega_0$ the
fermionic propagator is FL-like, and $\dg$ takes the FL form
discussed below). We then obtain to leading order in
$\omega_n/\omega_0$
\begin{equation}
\Delta \Gamma (\omega_n) = G(z) \ln\left(\frac{\omega_0}{\omega_n}
\right) \label{eq:App2DNFLbw}
\end{equation}
where $z$ is given in Eq.(\ref{eq:z}) and
\begin{eqnarray}
G(z)&=&\frac{\sqrt{3}}{4} \frac{z}{1+z^6} \left[ 1 + \frac{8}{3
\sqrt{3}}
z (z^4 -1) \right. \nonumber \\
& & \quad \left. + \sqrt{2} z^{3/2} (1-z^3) + \frac{2}{\pi} z^3 \ln
z \right] \label{eq:G}
\end{eqnarray}

In contrast, for the FL case, the integration of ${\hat q}$ in
Eq.(\ref{eq:dg_2d_bw5}) yields
\begin{equation}
 \dg(\omega_n) = -\frac{3}{16 \pi }\frac{1}{N_0 r}
 \int_{y_n}^{\infty} dy \frac{P(y)}{y}
 \label{eq:dg_2d_bw6}
\end{equation}
where $y=z^3 \hat{\nu}$ and
\begin{align}
 P(y) = &\frac{1}{1+y^2}
         \left[1+\frac{8}{3 \sqrt{3}} y^{1/3} (y^{4/3}-1)  \right. \nonumber \\
        &\left. + \sqrt{2} y^{1/2} (1-y) + \frac{1}{9\pi} y \ln y \right].
\end{align}
Performing the final integral over $y$, one obtains
\begin{equation}
 \dg(\omega_n) = -\frac{3}{16 \pi^2 }\frac{1}{N_0 r} R(y_n)
 \label{eq:dgfinal}
\end{equation}
where \begin{widetext}
\begin{align}
R(y_n)= & i \pi - \ln y_n + \frac{1}{2} \ln(1+y_n^2) - 2 \arctan y_n
\ln y_n  + \ln \left[ \frac{1+\sqrt{2} y_n^{1/6} -
y_n^{1/3}}{-1+\sqrt{2} y_n^{1/6} - y_n^{1/3}} \right] +
\frac{7}{3}\ln \left[ \frac{1+\sqrt{2-\sqrt{3}} y_n^{1/6} +
y_n^{1/3}}{1+\sqrt{2+\sqrt{3}} y_n^{1/6} + y_n^{1/3}} \right]
\nonumber \\
& + \frac{1}{3}\ln \left[ \frac{-1+\sqrt{2-\sqrt{3}} y_n^{1/6} -
y_n^{1/3}}{-1+\sqrt{2+\sqrt{3}} y_n^{1/6} - y_n^{1/3}} \right] -
\frac{i}{18 \pi} \left[ Li_2(i y_n) - Li_2 (-i y_n) \right]
\end{align}
\end{widetext}
and $Li_2$ is the dilogarithm function. Expanding $R(y_n)$ in the
limits $y_n \ll 1$ and $y_n \gg 1$, one obtains the results for
$\dg$ presented in Eqs.(\ref{eq:dg_2d_bw_lf}) and
(\ref{eq:dg_2d_bw_hf}), respectively.


\begin{thebibliography}{18}
\bibitem{com2}
 The result for $\dg$ in $D=2$, Eq.(\ref{eq:dg_2D}), can be further simplified
 by performing the integral over $\theta$ analytically via a standard
 integration over a closed circle in the complex plane. The resulting
 expression is then evaluated numerically.
\end{thebibliography}


\end{document}